\documentclass[prd,nofootinbib,reprint]{revtex4-1}
\pdfoutput=1

\usepackage[utf8]{inputenc}
\usepackage{hyperref}
\usepackage{amsmath}
\usepackage{amsfonts}
\usepackage{amssymb}
\usepackage{graphicx}
\usepackage{color}

\DeclareMathOperator\tr{tr}
\DeclareMathOperator\sign{sgn}
\DeclareMathOperator{\re}{Re}

\DeclareMathOperator{\diag}{diag}
\newcommand{\eq}[1]{Eq.~\eqref{#1}}

\newcommand{\hm}{\hat m}
\newcommand{\hmu}{\hat \mu}

\renewcommand{\epsilon}{\varepsilon}

\newcommand{\Ns}{N_s}
\newcommand{\Sdet}{\sigma}
\newcommand{\Y}{O}
\newcommand{\tPhi}{\Psi}
\newcommand{\tphi}{\psi}
\newcommand{\waux}{w_0}
\newcommand{\DOT}{.}
\newcommand{\tcobare}[1]{t^{#1}_{ij}}
\newcommand{\tco}[2]{\tcobare{#1}(#2;\theta)}
\newcommand{\tNmu}{\tco{N}{\mu}}
\newcommand{\tN}{\tco{N}{0}}
\newcommand{\ssum}{\sum_\Omega}
\newcommand{\Dphi}{\int_\phi }
\newcommand{\fac}{\gamma}

\newcommand{\one}{1}

\begin{document}	

\title{A subset solution to the sign problem in random matrix simulations}

\author{Jacques Bloch}
\affiliation{Institute for Theoretical Physics, University of Regensburg, 93040 Regensburg, Germany}

%\email{jacques.bloch@physik.uni-regensburg.de}
\date{May 24, 2012}
\revised{August 30, 2012}

\begin{abstract}
{
We present a solution to the sign problem in dynamical random matrix simulations of a two-matrix model at nonzero chemical potential. The sign problem, caused by the complex fermion determinants, is solved by gathering the matrices into subsets, whose sums of determinants are real and positive even though their cardinality only grows linearly with the matrix size. A detailed proof of this positivity theorem is given for an arbitrary number of fermion flavors. We performed importance sampling Monte Carlo simulations to compute the chiral condensate and the quark number density for varying chemical potential and volume. The statistical errors on the results only show a mild dependence on the matrix size and chemical potential, which confirms the absence of sign problem in the subset method. This strongly contrasts with the exponential growth of the statistical error in standard reweighting methods, which was also analyzed quantitatively using the subset method. Finally, we show how the method elegantly resolves the Silver Blaze puzzle in the microscopic limit of the matrix model, where it is equivalent to QCD.
}
\end{abstract}


\maketitle

\section{Introduction}

Our knowledge about QCD at finite baryon density is scarce, mainly because dynamical simulations at nonzero chemical potential are severely hindered by the sign problem, see Ref.~\cite{deForcrand:2010ys} for a review. The problem arises because the fermion determinant becomes complex and can no longer be included in the probabilistic weight in importance sampling methods. At small chemical potential the problem is mild and can be circumvented with methods, like reweighting, Taylor expansions and analytic continuation from imaginary to real chemical potential \cite{deForcrand:2010ys}. The cost of these methods typically grows exponentially with the volume and the chemical potential, such that they quickly become unusable.  For some specific models, tailor-made approaches were proposed which weaken or sometimes even solve the sign problem \cite{Chandrasekharan:2009wc,Wenger:2008tq,Aarts:2009uq,Bringoltz:2010iy,Mercado:2011ua}. 

Because of the equivalence between QCD in the $\varepsilon$-regime and the microscopic limit of chiral random matrix theory (RMT) \cite{Osborn:1998qb,Toublan:1999hx,Basile:2007ki}, useful spectral information about the Dirac operator in QCD can be gained from RMT, both at zero and nonzero chemical potential. RMT models for QCD at finite density where proposed by Stephanov \cite{Stephanov:1996ki} and Osborn \cite{Osborn:2004rf}.
These models not only allow the computation of the densities of complex eigenvalues of the Dirac operator \cite{Splittorff:2003cu,Osborn:2004rf,Akemann:2004dr}, which were successfully verified in quenched lattice simulations \cite{Akemann:2003wg,Bloch:2006cd,Akemann:2007yj}, but they also enable the study of the average phase of the fermion determinant \cite{Splittorff:2006fu,Splittorff:2007ck,Bloch:2008cf,Lombardo:2009aw,Bloch:2011jk}, which characterizes the severity of the sign problem in dynamical simulations. Moreover, the relation between the strongly oscillating spectrum, the sign problem and the chiral condensate at nonzero chemical potential was investigated, and a new mechanism was proposed to explain chiral symmetry breaking at finite chemical potential, which fundamentally differs from the Banks-Casher solution at zero chemical potential  \cite{Osborn:2005cy,*Osborn:2005ss,Osborn:2008jp}. At the same time this mechanism solves the Silver Blaze puzzle \cite{Cohen:2003kd,*Cohen:2004qp}, as subtle  cancellations ensure that thermodynamic observables are independent of the chemical potential below the nuclear matter threshold.

The aim of our work was to develop an importance sampling Monte Carlo method to simulate the dynamical partition function of random matrices, without using additional analytical input. One of the motivations is to learn from numerical solutions to the sign problem in simpler models to make progress in the solution of the sign problem in QCD.

We herein give a detailed report of a subset solution to the sign problem in dynamical simulations of the two-matrix model of Osborn at nonzero chemical potential \cite{Osborn:2004rf}, which we first proposed in an earlier letter \cite{Bloch:2011jx}. The principal feature of the method is the construction of subsets of matrices which have real and positive weights, but whose cardinality only grows linearly with the volume. This positivity was stated in the initial letter and its proof is given in full herein. 

The positive weights allow the use of importance sampling to construct a Markov chain of subsets, distributed according to the random matrix partition function. These subset samples were used to compute observables, like the chiral condensate and the average quark number density. The errors on the measurements clearly show that the new method is free of sign problem. This strongly contrasts with the results of standard reweighting methods, where the exponential blowup of the errors clearly signals the existence of the sign problem and leads to the failure of these methods. 

Interestingly, the statistical error on the average reweighting factors, which drives the sign problem in the reweighting methods, is a quantity that can also be computed using the subset method. This allowed for a quantitative determination of their exponential increase and for a comparison of various reweighting schemes. 

We further illustrate that the subset method remains efficient in the parameter region where large oscillations in the Dirac spectrum are crucial to get the correct value for the chiral condensate \cite{Osborn:2005cy,*Osborn:2005ss,Osborn:2008jp}.
Moreover, we show how the subset method resolves the Silver Blaze puzzle in the microscopic limit of the RMT model, where it is equivalent to QCD in the $\varepsilon$-regime.

Other approaches based on partial integrations or summations were considered in Refs.~\cite{Gocksch:1988iz,Kieu:1993gw,Anagnostopoulos:2001yb,Ambjorn:2002pz,Fodor:2007vv}.

This paper is composed as follows. In Sec~\ref{sec:rmt} we introduce the random matrix model. In Sec.~\ref{sec:subset} we present the subset method and give the positivity theorem, which allows for its use in Monte Carlo simulations. 
In Sec.~\ref{sec:results} we compare the numerical results for the chiral condensate and the quark number density obtained with the subset method and with standard reweighting methods, and explicitly compute the reweighting factors occurring in the latter.
In Sec.~\ref{sec:discussion} we discuss some implications of the positivity relation and show how the subset method resolves the Silver Blaze puzzle in the microscopic limit of RMT. Finally, we conclude in Sec.~\ref{sec:Conclusions}. In the appendix we give a proof of the positivity theorem, details on the numerical implementation, an overview of reweighting and some analytical RMT formulae. We also show how the Silver Blaze can be realized away from the microscopic limit.

\section{Random matrix model}
\label{sec:rmt}

In the two-matrix model of Osborn \cite{Osborn:2004rf} the complex matrices $\phi_1$ and $\phi_2$ of dimension $(N+\nu) \times N$ are distributed according to the unquenched partition function
\begin{align}
  Z = \int d\phi_1 d\phi_2 \, w(\phi_1) w(\phi_2) \, \prod_{f=1}^{N_f} {\det} \, D(\phi_1, \phi_2;\mu,m_f) ,
  \label{partfun} 
\end{align}
where the integration is performed over the real and imaginary parts of all matrix entries. The weights consist of a Gaussian part and a fermionic part originating from the $N_f$ dynamical quarks with masses $m_f$. Each random matrix has a Gaussian weight
\begin{align} 
  w(\phi_i) = (\fac N/\pi)^{N(N+\nu)} \exp(-\fac N \tr \phi_i^\dagger \phi_i) ,
\label{rmtdis}
\end{align}
where $\gamma \in \mathbb{R}^+$ and we will adopt the convention $\fac=1$ to conform with the standard discussions of this random matrix model. An alternative choice for $\fac$ will be briefly discussed in Sec.~\ref{sec:discussion}. Each fermion flavor contributes to the partition function with the determinant of its Dirac operator, which in this two-matrix model is given by
\begin{align} D(\phi_1, \phi_2;\mu,m)  =
  \begin{pmatrix}
    m & i\phi_1 + \mu\phi_2 \\
    i \phi_1^\dagger + \mu\phi_2^\dagger & m
  \end{pmatrix} ,
  \label{Ddef}
\end{align}
for a fermion of mass $m$ at chemical potential $\mu$ ($D$ has $\nu$ zero modes for $m=0$, where $\nu \geq 0$ without loss of generality).
In the presence of a chemical potential the fermion determinant becomes complex, such that the fermion weights can no longer be included in the probability distribution used in importance sampling Monte Carlo simulations. Standard solution methods used to circumvent this issue typically suffer from the sign problem, as the work needed to make reliable measurements on the statistical ensemble grows exponentially with the volume \cite{deForcrand:2010ys}. 

\section{Subset method}
\label{sec:subset}

Below we present a method, first introduced in Ref.~\cite{Bloch:2011jx}, which avoids the sign problem in dynamical simulations of random matrices. The main idea is to gather matrices into subsets which have net real and positive weights in the partition function \eqref{partfun}.

The subsets are constructed as follows: For any \textit{configuration} $\Phi=(\phi_1,\phi_2)$ we consider a set of configurations
\begin{align}
\Omega(\Phi)=\left\{ \tPhi(\Phi;\theta_n):  \theta_n = \frac{\pi n}{\Ns} \wedge n=0,\dotsc,\Ns\!-\!1 \right\} ,
\label{subset}
\end{align}
where $\tPhi(\Phi;\theta_n)=(\tphi_1, \tphi_2)$ are orthogonal rotations of the seed configuration $\Phi$ defined as
\begin{align}
\begin{pmatrix}
\tphi_1 \\
\tphi_2 
\end{pmatrix}
\equiv
\begin{pmatrix}
\hspace{2ex}\cos\theta & \sin\theta \\
 - \sin\theta & \cos\theta
\end{pmatrix}
\begin{pmatrix}
\phi_1 \\
\phi_2
\end{pmatrix} .
\label{tphi}
\end{align}
The initial motivation leading to the construction of the subsets using orthogonal rotations can be found in Ref.~\cite{Bloch:2011jx}.
The product $w(\tphi_1)w(\tphi_2)$ is independent of $\theta$ under the orthogonal rotation \eqref{tphi}, such that all the configurations in a subset $\Omega$ have the same Gaussian weight, which we denote as $W(\Omega)$.

The random matrix partition function \eqref{partfun} can then be rewritten as an equivalent integral over subsets $\Omega$ defined in \eqref{subset}, such that
\begin{align}
  Z = \int d\Omega \, W(\Omega) \, \Sdet_\Omega(\mu,m) , 
  \label{Zsubset}
\end{align}
where the fermionic subset weight, for $N_f$ degenerate quarks of mass $m$, is given by the sum of complex determinants,
\begin{align}
\Sdet_\Omega(\mu,m) = \sum_{n=0}^{\Ns-1} {\det}^{N_f}\! D(\tPhi_{n};\mu,m) ,
\label{sigma}
\end{align}
with $\tPhi_{n}\equiv\tPhi(\Phi;\theta_n) \in \Omega$.
The subset partition function \eqref{Zsubset} is equivalent to the random matrix partition function \eqref{partfun} as each configuration $\Phi=(\phi_1,\phi_2)$ of the random matrix ensemble can be used as seed of the corresponding subset $\Omega(\Phi)$, defined in Eq.~\eqref{subset}, and the set of all subsets forms an $\Ns$-fold covering of the original random matrix ensemble. 

The success of the method is based on the following \textit{positivity theorem:}  
For any $\Omega$ constructed according to \eqref{subset}, and for arbitrary $\mu<1$ and mass $m$, the fermionic subset weight $\Sdet_\Omega(\mu,m)$ defined in \eqref{sigma}, i.e., the sum of fermionic determinants of the $\Ns$ configurations $\tPhi(\Phi;\theta_n)$, is \textit{real} and \textit{positive} if $\Ns > N_f N$.

This theorem immediately follows from the identity
\begin{align}
\Sdet_\Omega(\mu,m) = (1-\mu^2)^{N_f(N+\nu/2)}\,\Sdet_\Omega\left(0,\frac{m}{\sqrt{1-\mu^2}}\right),
\label{conjectmneq0}
\end{align}
which holds for any $\mu$ and $m$ \textit{if} {$\Ns > N_f N$} and relates the fermionic subset weight at chemical potential $\mu$ and mass $m$ to the subset weight at zero chemical potential and effective mass $m_\mu=m/\sqrt{1-\mu^2}$. 
When looking at the right-hand side of Eq.~\eqref{conjectmneq0}, we observe that for zero chemical potential and mass $m_\mu \in \mathbb{R}$ the Dirac matrix generically has $\nu$ real eigenvalues $m_\mu$ and $N$ complex conjugate pairs $m_\mu \pm \imath \lambda$ with $\lambda\in\mathbb{R}^+$, such that its determinant is positive.
Hence, this right-hand side is real and positive for $\mu < 1$, and so too will be the subset weight $\Sdet_\Omega(\mu,m)$ on the left-hand side. For $\mu>1$ the effective mass becomes imaginary and the subset weights are no longer guaranteed to be of definite sign altogether. 
In the limit $m=0$ (and $\nu=0$ as the determinant is zero elsewise) the relation becomes 
\begin{align}
\Sdet_\Omega(\mu,0) = (1-\mu^2)^{N_fN}\Sdet_\Omega(0,0),
\label{conject}
\end{align}
for any $\mu$, which is the equation given in Ref.~\cite{Bloch:2011jx}. In that first publication an inequality was given for the case $m \neq 0$, which is now replaced by the general identity \eqref{conjectmneq0}. 

Note that for $\mu = 1$ Eq.~\eqref{conject} yields $\Sdet_{\Omega}(1,0)=0$, i.e., the sum of determinants exactly vanishes for $\mu=1$ and $m=0$. This corresponds to maximal non-hermiticity, where the average phase factor and the partition function are exactly zero, and the sign problem is maximal when using traditional solution methods. For nonzero mass, the limit of Eq.~\eqref{conjectmneq0} for $\mu \to 1$ is $\lim_{\mu\to 1} \Sdet_{\Omega}(\mu,m) = \Ns m^{N_f (2N+\nu)}$ and all the subsets in the partition function have identical fermionic weights, even though their Gaussian weights $W(\Omega)$ will vary.

The detailed proofs of the identities~\eqref{conjectmneq0} and \eqref{conject} are given in Appendix~\ref{sec:proof}.

The positivity implies that the subset weights $W(\Omega) \Sdet_\Omega(\mu,m)$ can be used to generate subsets of random matrices using importance sampling methods like the Metropolis algorithm. As $\Ns$ has to be larger than $N_f N$ to ensure positivity, it will be set to its optimal, i.e. smallest possible, value in the simulations, $\Ns=N_f N+1$. 

Using a sample of $N_\text{MC}$ subsets $\Omega_k$, $k=1\ldots N_\text{MC}$,  the expectation value of an observable $\Y$ in the RMT ensemble is approximated by the sample average
\begin{align}
\langle O \rangle \approx  \overline \Y &= \frac{1}{N_\text{MC}} \sum_{k=1}^{N_\text{MC}} \langle\Y\rangle_{\Omega_k} ,
\label{avgY}
\end{align}
where the \textit{subset measurement} is defined as
\begin{align}
\langle\Y\rangle_\Omega \equiv \frac{1}{\Sdet_{\Omega}} \sum_{n=0}^{\Ns-1} {\det}^{N_f}\! D(\tPhi_{n}) \, \Y(\tPhi_{n}) ,
\label{subsetavg}
\end{align}
with $\tPhi_{n} \in\Omega$ and $\langle 1 \rangle_\Omega = 1$, and we omit the arguments $\mu$ and $m$ from now on when no confusion is possible. 
The subset measurement is a modified subset average, which takes into account that the subsets are generated with a fermionic weight $\Sdet_{\Omega}$, while the individual matrices ought to be weighed by their respective Dirac determinants. 
In the simulations reported below, the subset weights and subset measurements were computed exhaustively, i.e., the determinants of \textit{all} $\Ns$ configurations in the subset were evaluated numerically and used in Eqs.~\eqref{sigma} and \eqref{subsetavg}.

\section{Numerical results}
\label{sec:results}

\subsection{Chiral condensate and quark number density}
\label{numsim}

We applied the subset method to compute the chiral condensate and the quark number density in the random matrix ensemble, which are given by
\begin{align}
\Sigma \equiv \frac{1}{2N}\frac{\partial\log Z}{\partial m} = \left\langle\frac{1}{2N} \tr D^{-1} \right\rangle
\label{eq:cc}
\end{align}
and
\begin{align}
n \equiv \frac{1}{2N}\frac{\partial\log Z}{\partial\mu} = \left\langle\frac{1}{2N} \tr \left[
\begin{pmatrix}
0 & \phi_2 \\
\phi_2^\dagger & 0
\end{pmatrix}
D^{-1}
\right] \right\rangle ,
\label{eq:qn}
\end{align}
respectively, following the derivation in Appendix~\ref{sec:implementation}. 

Using the block structure of the Dirac matrix \eqref{Ddef} more efficient formulae can be derived for the numerical evaluation of the fermion determinant, the chiral condensate and the quark number density, which are are given in Eqs.~\eqref{mdet}, \eqref{Simplem} and \eqref{nimplem}, respectively. 

\begin{figure*}
\includegraphics{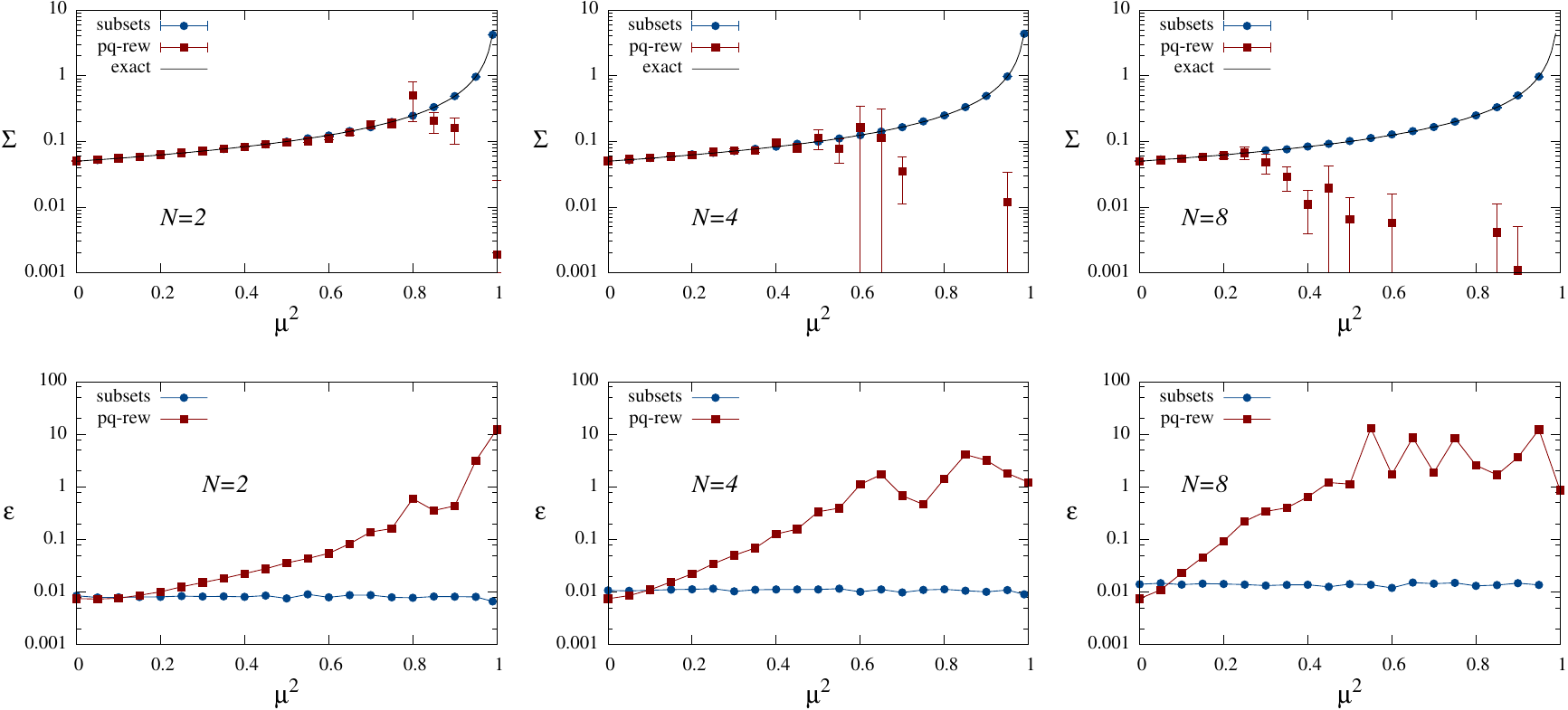}
\caption{Chiral condensate $\Sigma$ (top row) as a function of the chemical potential $\mu^2$ for the subset method and the phase-quenched reweighting method for $N=2,4,8$. The solid line shows the exact analytical result of Eq.~\eqref{Stheo}. The reweighting method fails for ever smaller $\mu^2$ when $N$ grows (some data points are negative and are left out of the semi-log plots). The corresponding relative statistical error $\varepsilon$ is shown in the bottom row. 
The error for the reweighting method grows very rapidly and should only be trusted as long as the method works (see top row).}
\label{fig-cc-vs-mu}
\end{figure*}

The simulations use the Metropolis algorithm to generate subsets $\Omega$ according to their statistical weights $W(\Omega)\,\Sdet_{\Omega}$. To compute the fermionic subset weights $\Sdet_{\Omega}$ the determinants of the $\Ns$ matrices in the subset are evaluated numerically and accumulated. It is important to emphasize that, even though the complex determinants fluctuate strongly when the chemical potential and volume increase, there is no sign problem in the computation of the individual subset weights $\Sdet_{\Omega}$, as all $\Ns$ contributing determinants are added and no statistical sampling is used. The subset weights are computed in a numerical\footnote{This contrasts with some other partial resummation/integration methods, where partial sums/integrals in the partition function are evaluated analytically, and only the integral over the remaining degrees of freedom is sampled numerically.}, but deterministic way.

The statistical sampling comes in when the successive subsets are generated in the Markov chain. This happens as follows: Start with a randomly chosen seed configuration and construct the corresponding set $\Omega_0$ using \eq{subset}. Assume now that the Markov chain has reached a subset $\Omega_t$ at Monte Carlo time $t$, then randomly choose a configuration in the current subset, generate a new configuration by making a random step on each matrix entry, construct the subset corresponding to this new seed configuration, and apply an accept-reject step to the newly proposed subset to generate the subset $\Omega_{t+1}$. This stepping procedure is repeated until we have generated a large enough sample to perform the desired measurement. 

In our simulations, we generated $N_\text{MC}=100,000$ subsets in each Markov chain, after equilibration was reached. 
The expectation value of an observable is evaluated by making a sample average of subset measurements, as prescribed in Eqs.~\eqref{avgY} and \eqref{subsetavg}. Each individual subset measurement is computed as a deterministic sum over $\Ns$ contributions. The quantities that fluctuate statistically during the Monte Carlo sampling are the subset measurements \eqref{subsetavg}. 
Successive measurements in the Markov chain are correlated and the number of independent measurements is smaller by a factor $2\tau$, where $\tau$ is the integrated autocorrelation time. The statistical errors on the measurements are determined using the standard error formula corrected for these autocorrelations. 

To compare the subset method with standard reweighting methods, described in Appendix~\ref{sec:rew},  the simulations were repeated using quenched, phase-quenched, $\mu$-quenched and sign-quenched reweighting, which are all expected to suffer from the sign problem \cite{Bloch:2011jk}. In reweighting methods observables are computed using the ratio \eqref{reweight}, where both numerator and denominator decrease exponentially with increasing volume.
The exponential increase of the work comes from the need to compute these exponentially small numbers from a statistical sampling of largely canceling contributions. 
The reweighting factors, i.e., the denominators in Eq.~\eqref{reweight}, will be discussed further in Sec.~\ref{sec:rewfac}.  
For the simulations with reweighting methods we used $N_\text{MC} \times \Ns $ random matrices in the Markov chains, such that the total number of generated matrices is the same as in the subset method. For the sake of clarity, we only show the results of phase-quenched reweighting in the figures below, as its results are representative for the various reweighting schemes.

We performed simulations with one dynamical fermion, i.e., $N_f=1$, of mass $m=0.1/2N$ and matrix sizes $N=2,\ldots,34$. The mass was chosen to be small with respect to the magnitude of the smallest eigenvalue, to ensure that dynamical effects are important. 

\begin{figure}
\includegraphics{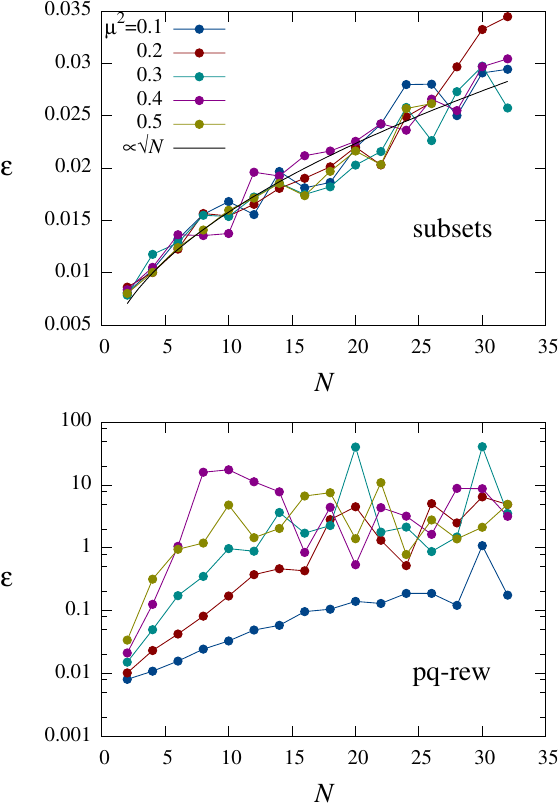}
\caption{Relative error $\varepsilon$ on the chiral condensate versus matrix size $N$ for various values of chemical potential $\mu^2=0.1,0.2,0.3,0.4,0.5$. The top plot shows the results of the subset method, for a fixed number of subsets. The full curve $\varepsilon(N) \propto \sqrt{N}$ serves to guide the eye. As a comparison, the bottom plot shows the relative error for phase-quenched reweighting on a semi-log plot (the color coding for $\mu^2$ is the same as in the top plot).}
\label{fig-cc-vs-N}
\end{figure}

\begin{figure*}
\includegraphics{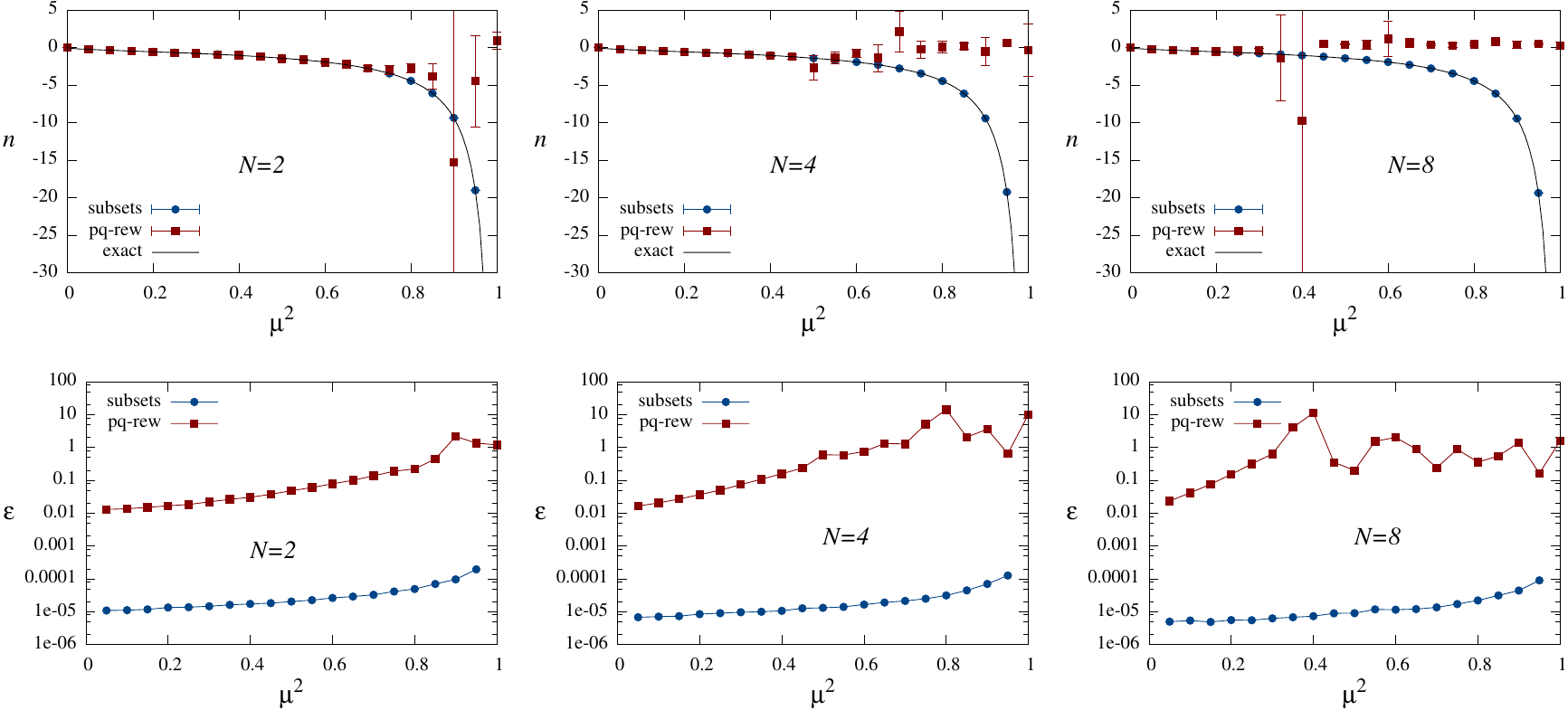}
\caption{Quark number density $n$ (top row) as a function of the chemical potential $\mu^2$ for the subset method and the phase-quenched reweighting method for $N=2,4,8$. The solid line shows the analytical result of Eq.~\eqref{ntheo}. Again, the reweighting method fails for ever smaller $\mu^2$ when $N$ grows. The corresponding relative statistical error $\varepsilon$ is shown in the bottom row. 
The error for the reweighting method grows very rapidly and should only be trusted as long as the method works (see top row).}
\label{fig:qn-vs-mu}
\end{figure*}

We measured the chiral condensate given by Eq.~\eqref{eq:cc}. These results were first presented in Ref.~\cite{Bloch:2011jx}. 
In Fig.~\ref{fig-cc-vs-mu} the condensate  is shown as a function of the chemical potential for matrices with sizes $N=2,4,8$.
We compare the results obtained using the subset method with those from phase-quenched reweighting and with the analytical results of Eq.~\eqref{Stheo}. The data, displayed in the top row, show that the reweighting method fails for smaller and smaller $\mu^2$ as the matrix size increases, due to the sign  problem. 
This strongly contrasts with the results of the subset method which are reliable up to much larger values of $\mu^2$ and agree with the analytical predictions. 

The corresponding relative statistical errors are shown in the bottom row.
At very small $\mu$, when the sign problem is not yet tangible, the error on the condensate is somewhat smaller for the reweighting method than for the subset method. This is easily explained by noting that at $\mu=0$ the determinants are all real and positive, such that importance sampling can be performed on the random matrices themselves. Sampling the partition function using subsets is then evidently somewhat less efficient. 
This feature persists for small, nonzero $\mu$, but very quickly the exponential growth of the error in the reweighting method, caused by the sign problem, makes the method unusable. At some value of $\mu$ the error estimate becomes meaningless, as the reweighting method completely fails. 
However, for the subset method, the relative accuracy of the measurements is nearly independent of the chemical potential, which confirms the absence of a sign problem and underscores the usefulness of the method.

We also studied how the relative statistical error on the chiral condensate varies as a function of the matrix size $N$ for fixed values of $\mu^2$, and show this $N$-dependence in Fig.~\ref{fig-cc-vs-N} for various values of $\mu^2$.
For a fixed number of subsets the error in the subset method (top panel) increases approximately as $\sqrt{N}$ and is independent of $\mu$ (the latter was already observed in Fig.~\ref{fig-cc-vs-mu}). 
If we fix the number of matrices, rather than the number of subsets, the error will increase with an additional factor $\sqrt{N}$ (as the subset size itself grows with $N+1$), such that the overall relative error will grow approximately linearly with $N$.
Conversely, to achieve a constant error the number of subsets would have to grow proportionally to $N$, i.e., the total number of matrices should approximately grow as $N^2$. The bottom plot shows the same quantity for phase-quenched reweighting (on semi-log scale). We observe that the error grows exponentially with $N$ until the reweighting method fails and the error is no longer reliable. Note that for both methods the additional cost for the numerical computation of the determinants is proportional to $N^3$.

\begin{figure}[b]
\includegraphics{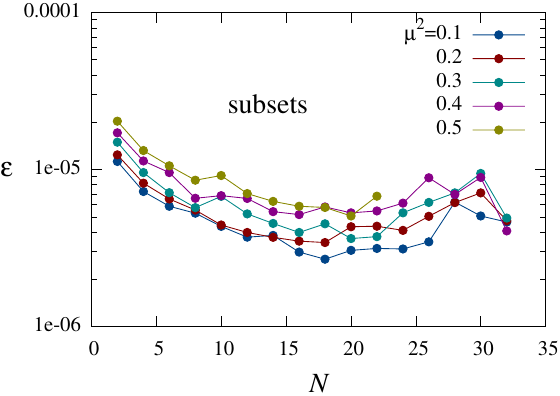}
\caption{Relative error $\varepsilon$ on the quark number density versus matrix size $N$ for various values of chemical potential $\mu^2=0.1,0.2,0.3,0.4,0.5$ using the subset method. }
\label{fig:qn-vs-N}
\end{figure}

Using the same Monte Carlo algorithm we also computed the quark number density given by Eq.~\eqref{eq:qn}. The variation of the number density and its relative statistical error as a function of the chemical potential are shown in Fig.~\ref{fig:qn-vs-mu}. The data in the top row confirm that the reweighting method badly suffers from the sign problem as $\mu$ and $N$ increase, while the subset method very nicely reproduces the analytical predictions of Eq.~\eqref{ntheo}.  
In the bottom row we observe that, for the reweighting method, the magnitude and variation of the error   are very similar to that on the chiral condensate and clearly signify a sign problem. However, for the subset method, the behavior of the statistical error not only confirms the absence of a sign problem,  but we observe that the relative error is much smaller than in the case of the chiral condensate. Moreover, even at small $\mu$ the error of the subset method is three orders of magnitude smaller than for the reweighting method, which is a very different behavior than for the chiral condensate. 
This surprising feature is due to the small variance of the quark number density over the Markov chain. This small variance can be understood from the results derived later in Sec.~\ref{sec:miclim}, where we show that the subset measurement \eqref{subsetavg} of the number density in this matrix model consists of a constant term, which is identical for all subsets, and a smaller term proportional to the chiral condensate; see Eq.~\eqref{qnrelsubset}. The first term does not contribute to the variance, such that the error on the number density is solely driven by the error on the latter, which explains why it is so small. From this we conclude that the quark number density in this model is especially well sampled by the importance sampling of subsets. 
The existence of subsets with large constant contribution to the quark number density is an interesting feature that could point to a possible search direction in other theories suffering from the sign problem.

We also measured the quark number density as a function of the matrix size $N$ using the subset method and show the relative error on these measurements in Fig.~\ref{fig:qn-vs-N}, for fixed number of sampled subsets. Here again, we only observe a mild dependence of the error on the matrix size, which confirms that there is no sign problem in the subset method.

\begin{figure}[b]
\includegraphics{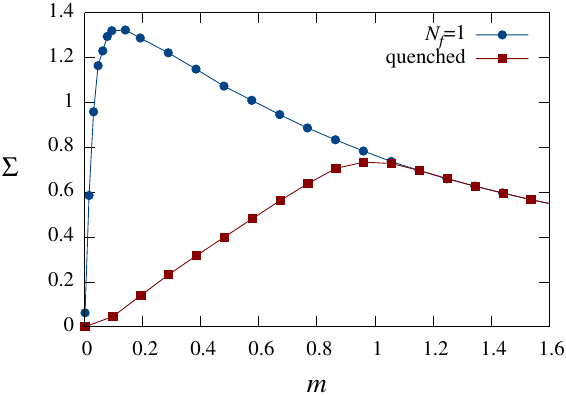}
\caption{Dynamical and quenched chiral condensate as a function of the quark mass for $\mu^2=0.6$ and $N=16$. The dynamical condensate was measured using the subset method.}
\label{fig:cc-vs-m}
\end{figure}

\begin{figure*}
\includegraphics[width=0.32\textwidth]{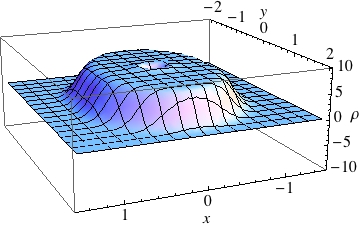}
\includegraphics[width=0.32\textwidth]{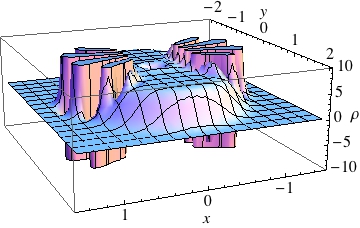}
\includegraphics[width=0.32\textwidth]{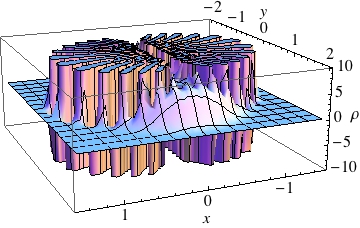}
\caption{Spectral density $\rho(x+\imath y)$ of the Dirac operator for $m=1.2,0.4,0.1$ (from left to right) with $\mu^2=0.6$ and $N=16$, computed using the formulae from Ref.~\cite{Osborn:2008jp}. For clarity, the height of the large oscillations was truncated in the figure.}
\label{Fig:spectrum}
\end{figure*}

\subsection{Spectral decomposition of the chiral condensate}

The spectral decomposition of the chiral condensate at nonzero chemical potential is quite remarkable, and the ability of the subset method to cope with this intricacy and reproduce the correct results is yet another test for its viability.

For this investigation we computed the variation of the chiral condensate as a function of the quark mass for a fixed chemical potential, both in the dynamical case ($N_f=1$), using the subset method, and in the quenched case. The results are shown in Fig.~\ref{fig:cc-vs-m}.  
As was shown in Refs.~\cite{Osborn:2005cy,Osborn:2005ss,Osborn:2008jp} the unquenched chiral condensate has a very different spectral decomposition depending whether the quark mass is outside or inside the cloud of complex eigenvalues of the Dirac operator, which has a width of about $2\mu^2$ \cite{Akemann:2004dr}. 
In Fig.~\ref{Fig:spectrum} we show typical spectral densities of the Dirac operator as the mass moves from outside to inside the eigenvalue spectrum.
When the mass is outside this cloud, the dynamical spectrum of the Dirac operator is almost identical to that of the quenched case (see leftmost panel of Fig.~\ref{Fig:spectrum}) and the dynamical chiral condensate should be very close to its quenched value. This is confirmed in Fig.~\ref{fig:cc-vs-m}, where we observe that both curves fall together when $m \gtrsim 2\mu^2$.
When the quark mass enters the cloud of eigenvalues the quenched chiral condensate steadily drops to zero. However, as can be seen in Fig.~\ref{fig:cc-vs-m}, the dynamical chiral condensate does not follow this trend and remains large all the way down to microscopically small masses. This peculiar behavior was explained in Ref.~\cite{Osborn:2008jp} by the very large oscillations which emerge in the unquenched eigenvalue spectrum at $z=\pm m$ when the mass enters the cloud of eigenvalues; see the spectral densities plotted in Fig.~\ref{Fig:spectrum}.
Subtle cancellations in these large spectral fluctuations compensate for the decline in the quenched contribution such that the dynamical chiral condensate remains large when the quark mass is inside the cloud of eigenvalues. 

Although these cancellations were computed analytically in Ref.~\cite{Osborn:2008jp}, it seemed unlikely that the dynamical chiral condensate could be determined to a good accuracy through numerical simulations, at least not in the region where the cancellations are important, as this is exactly where simulations are hampered by the sign problem \cite{Osborn:2005cy}.
Nevertheless, Fig.~\ref{fig:cc-vs-m} clearly shows that the subset method is able to compute the chiral condensate accurately, even in this critical parameter region.
The figure illustrates how the unquenched and quenched chiral condensates move apart when the mass enters the cloud of eigenvalues: The dynamical condensate remains large even though the quenched value steadily decreases. The efficiency of the subset method remains unaffected when the mass enters the cloud of eigenvalues and large spectral fluctuations are crucial to the determination of the chiral condensate. 
A similar mechanism relating the spectral oscillations of the baryon number Dirac operator to the quark number density was recently uncovered \cite{Ipsen:2012ug}.

\subsection{Reweighting factors}
\label{sec:rewfac}

Another quantity that is accessible to the subset method is the average reweighting factor occurring in the denominator of the reweighting formula \eqref{reweight}. 
These reweighting factors are important quantities in the study of the sign problem in reweighting methods as they decrease exponentially with increasing volume and chemical potential, and give rise to the exponentially growing error on the measurements.

The average reweighting factor for a target ensemble with complex weight $w$ simulated in an auxiliary ensemble $\waux$ is given by the expectation value $\left\langle {w/\waux} \right\rangle_{\!\waux}$, see Appendix \ref{sec:rew}. The direct computation of this expectation value in the auxiliary ensemble is obviously plagued by the sign problem, as it is precisely at the origin of the problem. However, this expectation value can be rewritten as
\begin{align}
r \equiv \left\langle \frac{w}{\waux} \right\rangle_{\!\waux} = \frac{\int dx \, w(x) }{\int dx \, \waux(x)}
= \left[\left\langle \frac{\waux}{w} \right\rangle_{w}\right]^{-1} ,
\label{rewfac}
\end{align}
which means that the reweighting factor in the auxiliary ensemble can be computed as an inverse expectation value in the unquenched ensemble. Although this cannot be evaluated with standard methods, as the unquenched ensemble has complex weights, it can easily be done using the subset method.

The unquenched expectation value $r^{-1}$ is then computed as the sample average \eqref{avgY} of subset measurements \eqref{subsetavg} of the inverse reweighting factor given by (assuming that the auxiliary weight only modifies the fermionic part of the action) 
\begin{align}
\left\langle \frac{\waux}{w} \right\rangle_{\Omega} = 
\frac{1}{\Sdet_{\Omega}(\mu,m)} \sum_{n=0}^{\Ns-1} w_0(\tPhi_n)
 .
 \label{rewfacsubset}
\end{align}
As the auxiliary weights $w_0$ are real and positive, so are the subset measurements \eqref{rewfacsubset}. Hence, the sample averages \eqref{avgY} do not involve any cancellations and the average reweighting factor \eqref{rewfac} can be efficiently determined by the subset method without encountering a sign problem, even though its value becomes exponentially small as the volume increases. 

\begin{table}[b]
\begin{tabular}{|c|c|c|}
\hline 
rew. scheme & ferm. part of $w_0$ & reweighting factor \\
\hline
quenched & 1 & $\det^{N_f}\!\! D(\mu)$ \\
phase-quenched & $R^{N_f}$ & $\exp(\imath N_f \varphi)$ \\
\hspace{1ex} $\mu$-quenched \hspace{1ex} & $\det^{N_f}\!\!D(0)$ & $\det^{N_f}\!\! D(\mu)/\det^{N_f}\!\! D(0)$\\
sign-quenched & $\left| \re \det^{N_f}\!\! D(\mu) \right|$ & $\sign\re\det^{N_f} \!\!D(\mu)$ \\
\hline
\end{tabular}
\caption{We list the four reweighting schemes studied in this paper. The auxiliary weights $w_0$ are products of the Gaussian weights and a fermionic part, given in the second column. The corresponding reweighting factor for each configuration is given in the third column. The fermion determinant is written as $\det D = R e^{\imath \varphi}$. 
}
\label{Table:rewschemes}
\end{table}

\begin{figure*}
\includegraphics{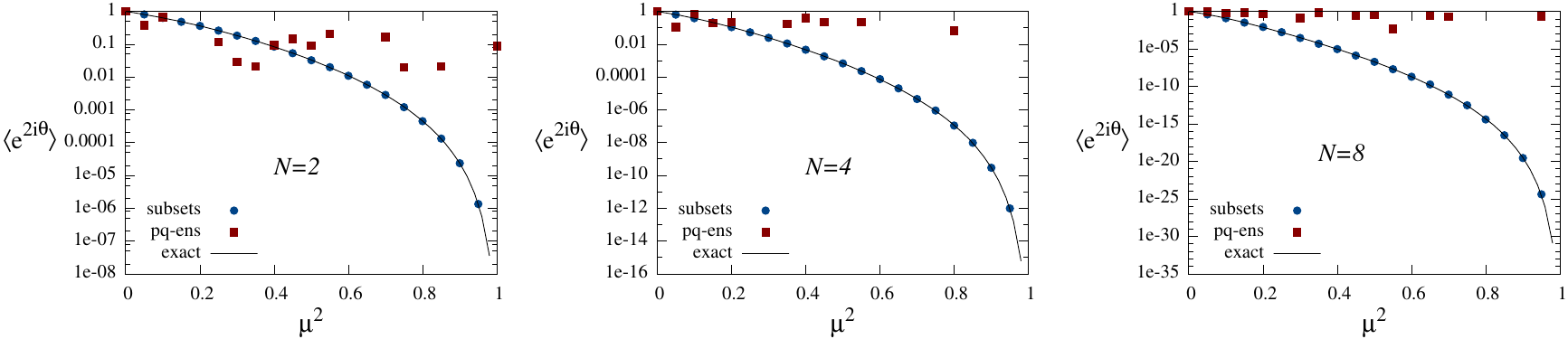}
\caption{Average two-fermion phase $\langle e^{2\imath\theta} \rangle_\text{pq}$ versus chemical potential in the $N_f=2$ phase-quenched ensemble for $m=0.1/2N$ and $N=2,4,8$. The results of the subset method (blue bullets) agree with the exact result of Eq.~\eqref{avgphase} (solid line), while the direct measurement in the phase-quenched ensemble (red squares) clearly suffers from the sign problem.}
\label{fig:phase-Nf=2-vs-mu}
\end{figure*}

\begin{figure*}
\includegraphics{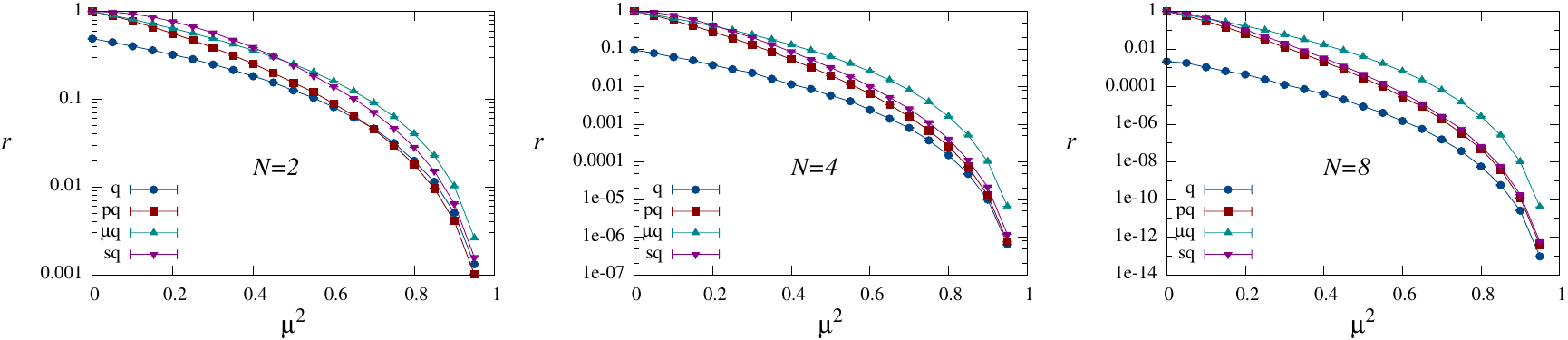}
\caption{Average reweighting factor $r$ versus chemical potential for the four reweighting schemes of Table \ref{Table:rewschemes} for $N_f=1$, $m=0.1/2N$ and $N=2,4,8$. The abbreviations in the legend are: (q) quenched, (pq) phase-quenched, ($\mu q$) $\mu$-quenched, (sq) sign-quenched. }
\label{fig:rewfac-vs-mu}
\end{figure*}

This enables us to compute and compare the reweighting factors for various reweighting schemes, 
which are summarized in Table~\ref{Table:rewschemes} and 
briefly described below:
\begin{itemize}

\item quenched: the configurations are generated by direct sampling of the Gaussian weights; the average reweighting factor is the average fermion determinant in the quenched ensemble,

\item phase-quenched: the auxiliary ensemble is generated using the magnitude of the complex determinants; the average reweighting factor is the average phase in that ensemble,

\item $\mu$-quenched or Glasgow scheme: the auxiliary ensemble is generated at zero chemical potential; 
the average reweighting factor is the average ratio of the determinants at $\mu$ and $\mu=0$ in that ensemble, 

\item sign-quenched: the configurations are generated according to the absolute value of the real part of the determinant; the average reweighting factor is the average sign of this real part in that ensemble. This reweighting scheme minimizes the relative variance of the reweighting factors \cite{deForcrand:2002pa}; see also \cite{Bloch:2011jk,Hsu:2010zza}.

\end{itemize}

The aim of this measurement was to compute the reweighting factors for these four different schemes using the subset method, as described in Eqs.~\eqref{avgY} and \eqref{rewfacsubset}, in order to verify and compare their exponential decrease when the chemical potential and volume are increased.

First, we verified the accuracy of the method for two-flavor ($N_f=2$) phase-quenched reweighting, by comparing the numerical data for the average reweighting factor with the analytical predictions given in Appendix~\ref{sec:averagephase}. The reweighting factor was computed using the subset method, as explained above, and also directly in the auxiliary ensemble. The agreement between the numerical and analytical data is illustrated in Fig.~\ref{fig:phase-Nf=2-vs-mu}. The very rapid decrease of the average phase factor is perfectly reproduced by the simulation data of the subset method, while the direct measurement in the phase-quenched ensemble fails because of the sign problem.

\begin{figure*}
\includegraphics{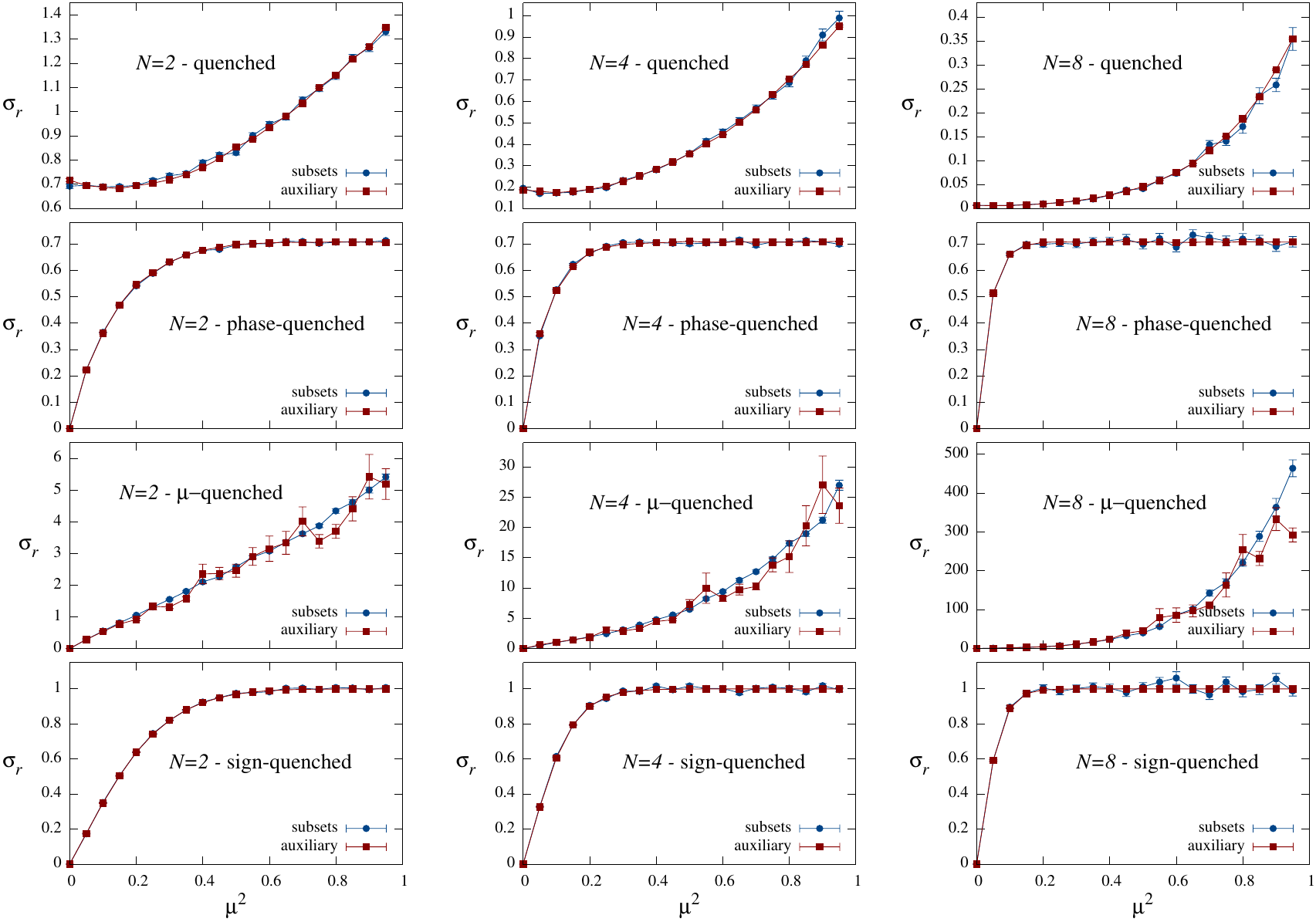}
\caption{Standard deviation $\sigma_r$ on the reweighting factor in various reweighting schemes as a function of the chemical potential for $N_f=1$, $m=0.1/2N$ and $N=2,4,8$, computed with the subset method (blue bullets) and directly in the auxiliary ensemble (red squares). From top to bottom we consider the quenched, phase-quenched, $\mu$-quenched and sign-quenched ensembles.}
\label{fig:sigma}
\end{figure*}

In general, the average reweighting factors in the various reweighting schemes cannot be computed analytically.
However, the subset method allows us to access these quantities numerically to good accuracy. 
We computed the average reweighting factors \eqref{rewfac} for the four schemes listed in Table~\ref{Table:rewschemes} for one flavor ($N_f=1$) and compare the results in Fig.~\ref{fig:rewfac-vs-mu}.

These data allow us to pinpoint the onset of the sign problem in the phase-quenched reweighting scheme by first locating the $\mu$-values where the reweighting method breaks down for the chiral condensate and quark number density in Figs.~\ref{fig-cc-vs-mu} and \ref{fig:qn-vs-mu}, and then reading off the corresponding reweighting factors for these $\mu$-values in Fig.~\ref{fig:rewfac-vs-mu}.
We find that the breakdown of the reweighting method, due to the sign problem, occurs when the average reweighting factor drops below $\approx 0.01$.

When comparing the four schemes in Fig.~\ref{fig:rewfac-vs-mu}, we observe that the Glasgow scheme has a somewhat larger reweighting factor than the other schemes.
Although this could naively be interpreted as hinting at a weaker sign problem, it is in reality only due to the fact that the magnitude of the determinants increases with increasing $\mu$, such that the values $w/w_0$ sampled in the Glasgow scheme are larger in magnitude than those in the other schemes. However, the sign problem is not actually caused by the size of the reweighting factor, but by its relative error, as it is the latter which propagates to every observable through Eq.~\eqref{reweight}. The reweighting factor itself is only an indicator for the sign problem, because an exponentially small value tells us that huge cancellations must take place.
To describe the exponential problem quantitatively one has to compute the relative error $\varepsilon_r$ on the average reweighting factor, 
\begin{align}
\varepsilon_r = \sqrt{\frac{2\tau}{N_\text{MC}}} \frac{\sigma_r}{r} ,
\label{stderr}
\end{align}
where 
\begin{align}
\sigma_r^2 = \left\langle \frac{(\re w)^2}{w_0^2} \right\rangle_{w_0} - r^2 
\label{var}
\end{align}
is the variance of the reweighting factor in the auxiliary ensemble. The variance involves the second moment of (the real part of) the reweighting factor, which 
can either be computed directly in the auxiliary ensemble, without encountering a sign problem, or can be computed using the subset method after rewriting it as:
\begin{align}
M_2 \equiv \left\langle \frac{(\re w)^2}{w_0^2} \right\rangle_{w_0}
&= r \left\langle \frac{\re w}{w_0} \right\rangle_{\re w} .
\label{secmom}
\end{align}
In Fig.~\ref{fig:sigma} we show the standard deviation $\sigma_r$ on the reweighting factor for the four reweighting schemes, where the second moment $M_2$ is either computed using the subset method or directly in the auxiliary ensemble (the reweighting factor $r$ is always computed with the subset method). We observe that the standard deviation is computed more accurately when $M_2$ is directly calculated in the auxiliary ensemble, except for the $\mu$-quenched ensemble where the subset method is much more accurate. Note that the latter ensemble is the only one where the standard deviation on the reweighting factor substantially grows with increasing chemical potential and volume.

\begin{figure*}
\includegraphics{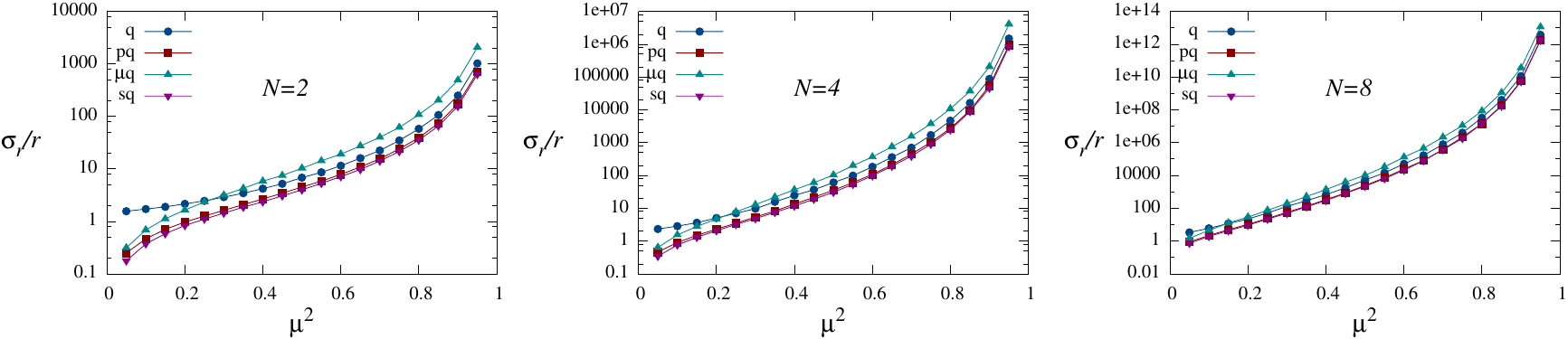}
\caption{Relative standard deviation $\sigma_r / r$ on the average reweighting factor for various reweighting schemes for $N_f=1$ and $m=0.1/2N$ versus chemical potential for $N=2,4,8$.}
\label{fig:rewerr-vs-mu}
\end{figure*}

We now merge the results of Figs.~\ref{fig:rewfac-vs-mu} and \ref{fig:sigma} to compute the  relative standard deviation $\sigma_r/r$, which is proportional to the standard error $\varepsilon_r$ of Eq.~\eqref{stderr}.
In Fig.~\ref{fig:rewerr-vs-mu} we compare $\sigma_r/r$ for the four different reweighting schemes.
The exponential problem becomes immediately clear, as the explosive growth of $\sigma_r/r$ with increasing $\mu$ and $N$ has to be compensated in simulations by sampling an exponentially large number of configurations to reach an acceptable accuracy for these reweighting schemes. The figure shows that, as expected, sign-quenched reweighting is doing slightly better than the other schemes \cite{deForcrand:2002pa}, even though it is closely followed by the phase-quenched scheme. For practical purposes the differences are however not really relevant as all reweighting schemes encounter the sign problem very early on. It is also interesting to note that the quenched reweighting scheme is not really outperformed, as we can generate uncorrelated random matrices by direct sampling of the Gaussian weights (this is specific to random matrices, as the matrix entries are distributed independently). For the other schemes we use the Metropolis algorithm to generate the configurations in the auxiliary ensemble such that autocorrelations have to be taken into account. The somewhat larger relative variance in the quenched scheme can then be compensated by the larger number of independent configurations generated for the same amount of work. On the other hand, the study shows that the Glasgow scheme performs worse than the other schemes. 

The quantitative analysis in this section confirmed that the sign problem is strong in all four reweighting schemes, and reliable numerical results can much better be obtained with the subset method, as seen in Sec.~\ref{numsim}.

\subsection{Remark on the Markov chain construction}

In our simulations we computed the fermionic subset weights $\Sdet_{\Omega}(\mu,m)$ by evaluating and adding up the $\Ns$ complex determinants at nonzero $\mu$. As an alternative, the Markov chain of subsets can be constructed by using the analytic formula \eqref{conjectmneq0} to compute the sampling weights $\Sdet_\Omega(\mu,m)$ from $\Sdet_\Omega(0,m_\mu)$. In this case one evaluates and sums up the fermionic determinants at $\mu=0$ and effective mass $m_\mu=m/\sqrt{1-\mu^2}$, and multiplies the sum with the suppression factor $(1-\mu^2)^{N_f(N+\nu/2)}$. This sum only involves positive real numbers such that no cancellations take place. The exponential smallness of the weights at large $\mu$ and $N$ comes entirely from the suppression factor, which is computed explicitly. As this factor is common to all subsets, it will drop out when taking ratios of probabilities in the Metropolis algorithm, such that it plays no role when generating the relevant subsets of the ensemble. 
This implies that the relevant subsets at chemical potential $\mu$ and mass $m$ are the same as those at $\mu=0$, albeit at a different, effective mass $m_\mu$. Moreover, the relevant subsets are independent of $\mu$ in the massless case.

Note, that this only concerns the Markov chain construction. The subset measurements \eqref{subsetavg} still require the determinants at the simulated $\mu$ and $m$, such that the simulation time will increase when using this alternative way to construct the Markov chain, as the $\Ns$ determinants have to be computed both at zero and nonzero chemical potential (the latter can be restricted to the independent subsets in the Markov chain if the autocorrelation times are known).

\section{Discussion}
\label{sec:discussion}

\subsection{Thermodynamic observables}
\label{sec:thermobs}

In this section we will see that \eq{conjectmneq0} allows us to derive some interesting relations for the thermodynamical observables.

For this, we first observe that the relation \eqref{conjectmneq0} for the subset weights percolates straightforwardly to the partition function, such that 
\begin{align}
Z_{N_f}(\mu;m) 
&= (1-\mu^2)^{N_f(N+\nu/2)} Z_{N_f}\left(0;m_\mu\right) .
\label{Zrel}
\end{align}
This relation agrees with the analytical expression for the partition functions derived using the method of orthogonal polynomials \cite{Osborn:2008jp,Akemann:2002vy}.

Using Eq.~\eqref{Zrel} one can relate the chiral condensate at $\mu \neq 0$ and $\mu=0$. The chiral condensate for $N_f=1$ is defined as
\begin{align}
\Sigma(\mu;m) &\equiv \frac{1}{2N} \frac{\partial}{\partial m} \log Z_{1}(\mu;m) ,
\label{ccdef}
\end{align}
and using Eq.~\eqref{Zrel} this can be rewritten as
\begin{align}
\Sigma(\mu;m) &= \frac{1}{2N} \frac{\partial}{\partial m} \log Z_{1}\left(0;m_\mu\right) \notag\\
&= \frac{1}{2N\sqrt{1-\mu^2}}\frac{\partial}{\partial m_\mu} \log Z_{1}\left(0;m_\mu\right) \notag\\
&= \frac{\Sigma\left(0;m_\mu\right)}{\sqrt{1-\mu^2}}
,
\label{sigmarel}
\end{align}
where we also used the chain rule and the definition of the effective mass $m_\mu$. 
This relation agrees with the analytical formula \eqref{Stheo} for the chiral condensate. 

A similar derivation can be performed for the average quark number density defined as (for $N_f=1$)
\begin{align}
n(\mu;m) &= \frac{1}{2N}\frac{\partial}{\partial\mu} \log Z_{1}(\mu;m).  \label{qndef}
\end{align}
This can be rewritten using Eq.~\eqref{Zrel} as
\begin{align}
n(\mu;m) &= -\left(1+\frac{\nu}{2N}\right)\frac{\mu}{1-\mu^2}  + \frac{1}{2N} \frac{\partial}{\partial\mu} \log Z_{1}\left(0;m_\mu\right) \notag\\
&= -\frac{\mu}{1-\mu^2} \left[ 1+\frac{\nu}{2N}  - \frac{m_\mu}{2N} \frac{\partial}{\partial m_\mu} \log Z_{1}\left(0;m_\mu\right) \right]\notag\\
&= -\frac{\mu}{1-\mu^2} \left[ 1+\frac{\nu}{2N}  - m_\mu \Sigma\left(0;m_\mu\right) \right] \notag\\
&= - \frac{\mu}{1-\mu^2} \left[ 1 +\frac{\nu}{2N} - m \Sigma\left(\mu;m\right) \right]  ,
\label{qnrel}
\end{align}
where we also used the definition of $m_\mu$, the chain rule, and Eqs.~\eqref{ccdef} and \eqref{sigmarel}. 
The quark number density can thus be written as a sum of its massless value and a correction term proportional to the quark condensate. 
This agrees with the analytical formula \eqref{ntheo} for the number density.

An interesting point is that the relations between the thermodynamic quantities at nonzero and zero chemical potential can also be derived at the subset level. For this, we note that for a thermodynamic observable $\langle \Y \rangle=$ $ \partial \log Z / \partial q$, with $q$ a fermionic parameter, \eq{Zsubset} yields
\begin{align}
  \langle \Y \rangle 
  &= \frac{1}{Z}\int d\Omega \, W(\Omega) \, \frac{\partial}{\partial q} \Sdet_\Omega(\mu,m) \notag\\
  &= \frac{1}{Z}\int d\Omega \, W(\Omega) \, \Sdet_\Omega(\mu,m)  \, \frac{\partial}{\partial q} \log \Sdet_\Omega(\mu,m) .
\end{align}
When the subsets are generated according to the sampling weights $W(\Omega) \, \Sdet_\Omega$, the subset measurements \eqref{subsetavg}, needed to compute the sample average \eq{avgY}, are now being given by
\begin{align}
\langle \Y \rangle_\Omega = \frac{\partial}{\partial q} \log \Sdet_\Omega(\mu,m) .
\label{subset_thermo}
\end{align}

We apply this formula to compute the individual subset contributions to the chiral condensate (for $N_f=1$), and find
\begin{align}
\Sigma_\Omega(\mu;m) &= \frac{1}{2N}\frac{\partial}{\partial m} \log \Sdet_\Omega(\mu,m) \notag\\
&= \frac{\Sigma_\Omega(0;m_\mu)}{\sqrt{1-\mu^2}}  ,
\label{sigmarelsubset}
\end{align}
where the derivation is analogous to that of \eq{sigmarel}.
Similarly we find that the contributions of the individual subsets to the quark number density (for $N_f=1$) are given by,
\begin{align}
n_\Omega(\mu;m) &= \frac{1}{2N}\frac{\partial}{\partial \mu} \log \Sdet_\Omega(\mu,m) \notag\\
&=  -\frac{\mu}{1-\mu^2} \left[1 +\frac{\nu}{2N} - m \Sigma_\Omega(\mu;m) \right] ,
\label{qnrelsubset}
\end{align}
where we followed the same steps as in \eq{qnrel}.
The subset relation \eqref{qnrelsubset} is important in the analysis of the statistical error in Fig.~\ref{fig:qn-vs-mu}. All subsets give a large common contribution to the number density, which leads to a small relative variance of the quark number density. Moreover, for $m=0$ all subsets give the same contribution to $n$, and the error on the measurement vanishes for the subset method, i.e., in the massless case a single subset would suffice to compute the correct number density.
Note that the large constant contribution to the quark number density is a non-trivial feature of the subset construction, which cannot be identified in the contributions of the individual random matrices.

\subsection{Microscopic limit and Silver Blaze puzzle}
\label{sec:miclim}

The equations derived above also show how the Silver Blaze puzzle\footnote{The \textit{Silver Blaze} puzzle in QCD refers to the fact that, at zero temperature and for a chemical potential less than approximately one third of the nucleon mass, the free energy and the thermodynamical observables are independent of $\mu$ \cite{Cohen:2003kd,*Cohen:2004qp}.} is resolved in the subset method. To be equivalent to QCD the microscopic limit of RMT has to be considered, where $\hat m = 2Nm$ and $\hat\mu^2=2N\mu^2$ are kept fixed when $N \to \infty$. In this limit Eqs.~\eqref{sigmarel} and \eqref{qnrel} lead to
\begin{align}
\hat \Sigma(\hat\mu;\hat m) = \hat \Sigma\left(0;\hat m \right) \quad \text{and} \quad
\hat n (\hat\mu;\hat m) = 0 ,
\end{align}
where we introduced the microscopic limits
\begin{align}
\hat f(\hat\mu;\hat m) = \lim_{N\to\infty} f(\hat\mu/\sqrt{2N}; \hat m/2N) . 
\end{align}
The chiral condensate and the quark number density are thus independent of the chemical potential in the microscopic limit of RMT.
Even more, following Eqs.~\eqref{sigmarelsubset} and \eqref{qnrelsubset} we observe that the contribution from each individual subset to the thermodynamic quantities is independent of $\mu$ in the microscopic limit, such that the Silver Blaze puzzle is in fact already resolved at the subset level.

Note that the prefactor in \eqref{conjectmneq0} generates an exponential factor in the microscopic limit of the RMT partition function, as
\begin{align}
\lim_{N\to\infty} (1-\mu^2)^{N_f N} &= \lim_{N\to\infty} \left(1-\frac{\hmu^2}{2N}\right)^{N_f N} \notag\\
&= \exp\left(  -\frac{N_f \hmu^2}{2} \right) .
\label{expsup}
\end{align}
Interestingly, this exponential factor is already generated within each subset individually as it originates from \eqref{conjectmneq0}. 
In the RMT model the free energy density is defined as
\begin{align}
F(\mu;m) = -\frac{1}{2N} \log Z(\mu;m),
\end{align}
which also becomes independent of $\mu$ in the microscopic limit because
\begin{align}
\hat F(\hmu;\hm) = \lim_{N\to\infty} \frac{N_f \hmu^2}{4N} + \hat F(0;\hm) = \hat F(0;\hm),
\end{align}
where we used Eqs.~\eqref{Zrel} and \eqref{expsup}. Although the factor \eqref{expsup} does not occur in the partition function of chiral perturbation theory and is an artifact of the random matrix model, it is not relevant when discussing its universal properties, as it leaves the microscopic eigenvalue correlations unchanged \cite{Halasz:1999gc,Ipsen:2012ug}. 

\begin{figure}
\includegraphics{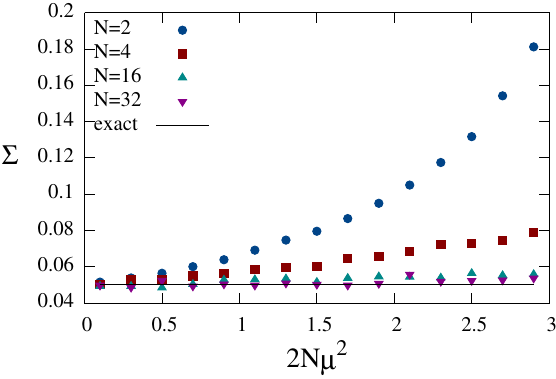}
\caption{Chiral condensate $\Sigma$ versus microscopic chemical potential $\hat\mu^2$ for $\hm=0.1$ and increasing values of $N$. As $N$ gets larger the numerical results converge towards the analytical microscopic value of Eq.~\eqref{theomiclim}. }
\label{fig:miclim}
\end{figure}
The subset method also allows a numerical investigation of the convergence towards the microscopic limit, as $N \to\infty$ with fixed microscopic parameters $\hmu$ and $\hm$. This is illustrated in Fig.~\ref{fig:miclim} where we plot the chiral condensate as a function of the microscopic chemical potential $\hat\mu^2$ for different values of $N$. As $N$ increases the chiral condensate converges to its microscopic limit given by Eq.~\eqref{theomiclim}. 

It is interesting to note that choosing $\fac=1-\mu^2$ (assuming $\mu\leq 1$) from the onset in the Gaussian weights \eqref{rmtdis} not only gets rid of the spurious factor \eqref{expsup}, but makes the partition function independent of the chemical potential.  This was proven using the method of orthogonal polynomials in Ref.\ \cite{Osborn:2004rf}, but it also directly follows from the subset relation \eqref{conjectmneq0} as is shown in Appendix \ref{app:gamma}. This modification has the salient feature that the Silver Blaze is then satisfied for any $N$, even away from the microscopic limit. The $\mu$-independence of $Z$ does not, however, alleviate the sign problem of the model, as the fermion determinants still exhibit huge fluctuations when the quark mass enters the cloud of eigenvalues. In fact, we know that these large oscillations are essential to resolve the Silver Blaze puzzle at large chemical potential. The efficacy of the subset method to solve the sign problem remains intact as this choice of $\fac$ merely cancels the prefactor in \eqref{Zrel} and rescales the fermion mass (see Eq.~\eqref{Zgamma}). The results for this $\fac$ can easily be related to those previously computed with $\fac=1$, and therefore we will not present explicit numerical results for this alternative choice. One observation is that the constant contribution to the quark number density \eqref{qnrelsubset}, which becomes large far from the microscopic limit, will cancel out in this case. However, even though their results differ at finite $N$, both values of $\gamma$ yield the same universal limit.

\section{Conclusions}
\label{sec:Conclusions}

In this paper we have presented a solution to the sign problem in dynamical simulations of the two-matrix model of Osborn at finite chemical potential.
The random matrices are gathered into subsets, which have real and positive fermionic weights while their cardinality only grows linearly with the matrix size. 
A detailed proof of the positivity theorem for the subset weights was given. 

The positive subset weights make it possible to sample the partition function with an importance sampling Monte Carlo method and generate a Markov chain of relevant subsets. 
As the chemical potential and the matrix size increase, the weights of the subsets rapidly decrease, but  without causing a sign problem in the simulations. In contrast to standard reweighting methods the large cancellations, inherent to simulations at real chemical potential, do not happen through statistical sampling of the ensemble but are confined inside the subsets, where the weights and measurements are computed in a deterministic way from a small number of contributions. The ensuing subset measurements, which are used to compute the sample averages, do not suffer from large statistical fluctuations so that the standard error on the simulation result is well under control.

The method was used to compute the chiral condensate and quark number density accurately over a large parameter range, showing that the subset method has no sign problem, even in regions where the reweighting methods are unusable. The method is especially well suited to compute the quark number density in this model, as the statistical error on this quantity is extremely small, which was understood from analytical considerations. 

The subset method also enabled us to compute the reweighting factors, appearing in the standard reweighting methods, and their relative standard errors. This explicitly revealed the exponential increase in the work required by these reweighting methods, signaling the presence of the sign problem. 

We also showed how the positivity relation resolves the Silver Blaze puzzle in the microscopic limit of the random matrix model, where it is equivalent with QCD, and how this mechanism already works at the subset level.

The important question whether the subset method can be applied to physical systems and ultimately to QCD itself has not yet been answered and is the focus of current research.

\begin{acknowledgments}
I would like to thank Falk Bruckmann, Philippe de Forcrand and Tilo Wettig for useful discussions. This work was supported by the DFG collaborative research center SFB/TR--55. 
\end{acknowledgments}

\appendix

\section{Proof of the positivity theorem}
\label{sec:proof}

In this appendix we prove the positivity theorem for the subset weights, which is at the basis of the subset method. We first prove the massless relation \eqref{conject} for a single fermion, before extending it to an arbitrary number of $N_f$ massless fermions. Consequently we generalize the identity to the massive case and prove relation \eqref{conjectmneq0}, first for one massive fermion and finally for $N_f$ fermions with degenerate mass $m$.

\subsection{One massless fermion}
\label{sec:conject_m=0}

Using its block structure, the determinant of the Dirac matrix \eqref{Ddef} for a massless fermion, with $\nu=0$ or neglecting the zero modes, is given by (see Appendix~\ref{sec:implementation})
\begin{align}
\det D_\theta(\mu) = \det Q_\theta(\mu),
\label{detQ}
\end{align}
where we introduce, for brevity, $D_\theta \equiv D(\tPhi(\Phi;\theta))$ for a rotated configuration $\tPhi(\Phi;\theta)=(\tphi_1,\tphi_2)$ defined in Eq.~\eqref{tphi}, and $Q_\theta = -B A$ is an $N \times N$ matrix with 
\begin{align}
\begin{cases}
A = \imath\psi_1 + \mu\psi_2 \\
B = \imath\psi_1^\dagger + \mu\psi_2^\dagger
\end{cases}
 .
\label{AB}
\end{align}
When expanding the product in $Q_\theta$ we find
\begin{align}
Q_\theta(\mu) &= \psi_1^\dagger\psi_1 - \mu^2\psi_2^\dagger\psi_2
- \imath\mu(\psi_1^\dagger\psi_2+\psi_2^\dagger\psi_1) \;,
\end{align}
and after substituting the definition \eqref{tphi} this becomes
\begin{align}
Q_\theta(\mu)
&= 
 \left( \cos^2\theta\; a
+ \sin^2\theta\; b
+\sin\theta\cos\theta\; c \right) \notag\\
& - \mu^2 \left(\sin^2\theta\; a 
+ \cos^2\theta\; b 
-\sin\theta\cos\theta\; c \right) 
\label{Dtheta}\\
&- \imath\mu \left( (\cos^2\theta-\sin^2\theta) c
+ 2\sin\theta\cos\theta \; (b - a) \right) \:, \notag
\end{align}
with the $N\times N$ matrices $a=\phi_1^\dagger\phi_1$, $b=\phi_2^\dagger\phi_2$ and $c=\phi_1^\dagger\phi_2+\phi_2^\dagger\phi_1$. 
After gathering the terms in $a$, $b$ and $c$ this can be rewritten as
\begin{align}
Q_\theta(\mu) &= f_a \, a + f_b \, b + f_c \, c  \;,
 \label{gatherabc}
\end{align}
with 
\begin{align}
\left\{
\begin{aligned}
f_a &= \cos^2\theta - \mu^2\sin^2\theta + 2\imath\mu \cos\theta\sin\theta \\
f_b &= \sin^2\theta - \mu^2 \cos^2\theta - 2\imath\mu \cos\theta\sin\theta \\
f_c &= (1+\mu^2)\sin\theta \cos\theta -  \imath\mu (\cos^2\theta-\sin^2\theta) 
\end{aligned}
\right. ,
\end{align}
which, interestingly, can be further simplified to
\begin{align}
\left\{
\begin{aligned}
f_a &= (\cos\theta + \imath\mu \sin\theta)^2 \\
f_b &= (\sin\theta - \imath\mu \cos\theta)^2\\
f_c &= (\cos\theta + \imath\mu \sin\theta)(\sin\theta - \imath\mu \cos\theta)
\end{aligned}
\right. .
\label{fabc}
\end{align}

Using the Leibniz formula for determinants, Eq.~\eqref{detQ} can be written as
\begin{align}
\det D_\theta(\mu) = \hspace{-1ex}\sum_{i_1,i_2,\cdots,i_N=1}^N \hspace{-3ex}\epsilon_{i_1,i_2,\cdots,\i_N} 
Q_{1 i_1} Q_{2 i_2} \cdots Q_{N i_N} ,
\label{Leibniz}
\end{align}
where $\epsilon_{i_1,i_2,\cdots,\i_N}$ is the antisymmetric Levi-Civita symbol and $Q_{ij}$ are the entries of $Q_\theta$. 
Each term in the sum is a product of $N$ matrix components, which, according to Eq.~\eqref{gatherabc}, can be written as $Q_{ij} = f_a a_{ij} + f_b b_{ij} + f_c c_{ij}$. The coefficients $f_a$, $f_b$, and $f_c$, given in \eqref{fabc}, are functions of $\mu$ and $\theta$, which are independent of the indices $i$ and $j$. After expanding all the products in \eqref{Leibniz} and gathering terms with equal powers of $f_a$, $f_b$ and $f_c$, the determinant can be written as 
\begin{align}
\det D_\theta(\mu) = \sum_{i,j=0}^N \tNmu \, H_{ij}^N(a,b,c) ,
\label{tabc}
\end{align}
where 
\begin{align}
\tNmu \equiv f_a^i f_b^j f_c^{N-i-j}
\label{t}
\end{align}
and $H_{ij}^N(a,b,c)$ is a sum of signed products, each containing $i$, $j$ and $N-i-j$ components of the matrices $a$, $b$ and $c$, respectively, which is implicitly defined by identifying Eqs.~\eqref{Leibniz} and \eqref{tabc}. The $\mu$ and $\theta$ dependence of the determinant \eqref{tabc} is completely contained in the coefficients $\tcobare{N}$ defined in Eq.~\eqref{t}. Using Eq.~\eqref{fabc} these coefficients can be simplified to 
\begin{align}
\tNmu &= (\sin\theta - \imath\mu \cos\theta)^\alpha (\cos\theta + \imath\mu \sin\theta )^\beta ,
\label{tij}
\end{align}
where $\alpha=N-i+j$ and $\beta=N+i-j$, and thus $\alpha+\beta=2N$ with $0 \leq \alpha,\beta \leq 2N$.

To prove the conjecture \eqref{conject} it is sufficient to prove that the identity holds term by term in the formula \eqref{tabc}, i.e,
\begin{align}
\ssum \tNmu = (1-\mu^2)^N \ssum \tN  ,
\label{conjecture}
\end{align}
for all $0 \leq i, j \leq N$, where we denoted the subset summation as
\begin{align}
\ssum f(\theta) \equiv \sum_{n=0}^{\Ns-1} f(n \pi/\Ns) ,
\label{subsetsum}
\end{align}
with the subset $\Omega$ defined in Eq.~\eqref{subset}.

To prove Eq.~\eqref{conjecture} we first convert the trigonometric functions in Eq.~\eqref{tij} into complex exponentials using
\begin{align}
\cos\theta = \frac{e^{\imath\theta}+e^{-\imath\theta}}{2} , \quad
\sin\theta = \frac{e^{\imath\theta}-e^{-\imath\theta}}{2\imath} , 
\end{align}
which leads to
\begin{align}
\tNmu 
&= \frac{(-\imath)^\alpha}{2^{2N}} 
\left[ (1+\mu) e^{\imath\theta} - (1-\mu) e^{-\imath\theta} \right]^\alpha \notag\\
& \times \left[ (1+\mu) e^{\imath\theta} + (1-\mu) e^{-\imath\theta} \right]^\beta .
\end{align}
We apply the binomial formula to expand both powers and find
\begin{align}
\tNmu 
&= \frac{\imath^\alpha}{2^{2N}} 
\sum_{k=0}^\alpha \sum_{\ell=0}^{\beta} 
(-)^{k} \binom{\alpha}{k} \binom{\beta}{\ell}
(1+\mu)^{k+\ell} \notag\\
&\times (1-\mu)^{2N-k-\ell} \,
e^{2\imath (k+\ell-N) \theta} \,,
\label{tijbinom}
\end{align}
where we also used that $\alpha+\beta=2N$.
The $\theta$ dependence of Eq.~\eqref{tijbinom} is completely contained in the exponential function and its subset 
sum \eqref{subsetsum} is 
\begin{align}
S_\Omega \equiv \ssum e^{2\imath q \theta} = \sum_{n=0}^{\Ns-1} e^{2\pi\imath  \frac{q n}{\Ns}} ,
\end{align}
where $q\equiv k+\ell-N$ is an integer $\in [-N,N]$.
$S_\Omega$ is a sum over the $q$-th powers of all the $\Ns$-th roots of unity. It can be computed by writing it as a geometric series,
\begin{align}
S_\Omega = \sum_{n=0}^{\Ns-1} \left(e^{2\pi\imath  \frac{q}{\Ns}}\right)^n .
\label{SOmega}
\end{align}
We distinguish two cases, depending whether $q/\Ns$ is integer or not.
For $q/\Ns \notin \mathbb{Z}$ we have $e^{2\pi\imath q/\Ns} \neq 1$, and the sum of the geometric series is
\begin{align}
S_\Omega = \frac{1-e^{2\pi\imath q}}{1-e^{2\pi\imath  \frac{q}{\Ns}}} = 0,
\end{align}
which vanishes as $q$ is integer by definition. If $q/\Ns \in \mathbb{Z}$ then $e^{2\pi\imath q/\Ns} = 1$ and the sum \eqref{SOmega} can be computed explicitly,
\begin{align}
S_\Omega = \Ns .
\end{align}

If we take \underline{$\Ns > N$}, the ratio $q/N_s$ will be non-integer and $S_\Omega$ zero for all $q$, except for $q=0$, as $q$ is an integer varying from $-N$ to $N$.
Hence, when summing Eq.~\eqref{tijbinom} over the subset, only terms for which $k+\ell=N$ will survive. After considering both cases $\alpha \gtrless \beta$, the subset sum can be written as
\begin{align}
\ssum \tNmu &=  (1-\mu^2)^{N} \omega_{ij} ,
\label{I9}
\end{align}
where
\begin{align}
\omega_{ij} &=
 \frac{\Ns}{2^{2N}} (\imath\sign\Delta)^{N^-} \sum_{k=0}^{N^-}
(-)^{k} \binom{N^-}{k} \binom{N^+}{N-k}
\label{omega}
\end{align}
with $N^\pm=N\pm|\Delta|$ and $\Delta=i-j$. 

As $\omega_{ij}$ is independent of $\mu$, \eq{I9} immediately implies that
\begin{align}
\ssum \tNmu 
&= (1-\mu^2)^{N} \ssum \tN .
\label{Tij}
\end{align}
Because this relation holds for any $i$ and $j$ in Eq.~\eqref{tabc}, it also holds for the sum over these indices, which proves the conjecture \eqref{conject} for $N_f=1$, i.e.,
\begin{align}
\ssum \det D_\theta(\mu)
&= (1-\mu^2)^{N} \ssum \det D_\theta(0) .
\end{align}

\subsection{$\mathbf{N_f}$ massless fermions}
\label{sec:Nfneq1}

To extend the theorem to $N_f > 1$, the determinant \eqref{tabc} has to be multiplied $N_f$ times so that many more terms are generated. Nevertheless, after exponentiation the global structure of the fermionic weight remains similar to that of Eq.~\eqref{tabc} and can be written as the following linear combination
\begin{align}
{\det}^{N_f} D_\theta(\mu) = \sum_{i,j=0}^{N_f N} \tco{N_f N}{\mu} \, \hat H_{ij}^{N_f N}(a,b,c) ,
\label{tabcNf}
\end{align}
where $\tcobare{N_f N}$ is defined in Eq.~\eqref{t} and the implicitly defined function $\hat H$ only depends on the components of the matrices $a$, $b$ and $c$, which were defined right after \eq{Dtheta}.
The remainder of the proof is identical to that for $N_f=1$ with $N$ replaced by $N_fN$, which eventually leads to
\begin{align}
\ssum \tco{N_f N}{\mu} 
&= (1-\mu^2)^{N N_f} \ssum \tco{N_f N}{0} .
\end{align}
Together with Eq.~\eqref{tabcNf} this proves the conjecture \eqref{conject} for arbitrary $N_f$:
\begin{align}
\ssum {\det}^{N_f} D_\theta(\mu)
&= (1-\mu^2)^{N_f N} \ssum {\det}^{N_f} D_\theta(0) .
\end{align}

\subsection{One massive fermion}

For the massive case we use the determinant block formula \eqref{mdet},
\begin{align}
\det D_\theta(\mu,m) = m^{\nu} \det Q_\theta(\mu,m) ,
\label{detDmum}
\end{align}
with $Q_\theta = m^2 - BA$, and $A$ and $B$ defined in Eq.~\eqref{AB}. To prove the conjecture \eqref{conjectmneq0} for $N_f=1$ we first study $\det Q_\theta$. If we repeat the steps leading to Eq.~\eqref{gatherabc} we now find
\begin{align}
Q_\theta(\mu,m) = m^2 + f_a a + f_b b + f_c c ,
\end{align}
where $f_a$, $f_b$ and $f_c$ are defined in \eqref{fabc}.
After using the Leibniz formula \eqref{Leibniz} and expanding all the products, as we did before when deriving Eq.~\eqref{tabc} for the massless case, we now find
\begin{align}
\det Q_\theta(\mu,m) = \sum_{h,i,j=0}^N \hspace{-1ex} \tco{N-h}{\mu} \, m^{2h} \tilde H_{i,j}^{N-h}(a,b,c),
\label{A27}
\end{align}
where $\tcobare{N-h}$ is defined in \eqref{t} and contains the full $\theta$ and $\mu$ dependence, while $\tilde H$ is implicitly defined by \eqref{A27} and depends on the components of the matrices $a$, $b$ and $c$.
We can repeat the whole argument of Sec.~\ref{sec:conject_m=0} on the subset sum of the $t$-coefficients, after replacing $N$ by $N-h$. In analogy to \eqref{Tij} this leads to
\begin{align}
\ssum \tco{N-h}{\mu} 
&= (1-\mu^2)^{N-h} \ssum \tco{N-h}{0} .
\label{tinter}
\end{align}
This identity still depends on the summation index $h$, and after substitution in the subset sum of \eqref{A27} we find
\begin{align}
\lefteqn{\ssum \det Q_\theta(\mu,m)} \notag\\
&= \ssum \sum_{h,i,j=0}^N \hspace{-1ex} \tco{N-h}{\mu} \, m^{2h} \tilde H_{i,j}^{N-h}(a,b,c) \notag\\
&= (1-\mu^2)^{N} \ssum \sum_{h,i,j=0}^N \hspace{-1ex} \tco{N-h}{0} \frac{m^{2h}}{(1-\mu^2)^{h}} \tilde H_{i,j}^{N-h}(a,b,c) \notag\\
&= (1-\mu^2)^{N} \ssum \det Q_\theta\left(0,\frac{m}{\sqrt{1-\mu^2}}\right) ,
\label{conjectmassive}
\end{align}
where the last equation is easily derived by inspection, after setting $\mu=0$ and replacing $m$ by $m/\sqrt{1-\mu^2}$ in Eq.~\eqref{A27}. 
After multiplying \eq{conjectmassive} with $m^\nu$ and using Eq.~\eqref{detDmum}, we find
\begin{align}
&\ssum \det D_\theta(\mu,m) \notag\\
&= (1-\mu^2)^{N} m^\nu \ssum \det Q_\theta\left(0,\frac{m}{\sqrt{1-\mu^2}}\right) \notag\\
&= (1-\mu^2)^{N+\nu/2} \ssum \det D_\theta\left(0,\frac{m}{\sqrt{1-\mu^2}}\right) .
\end{align}
This proves the conjecture \eqref{conjectmneq0} for $N_f=1$:
In the massive case the fermionic subset weight at chemical potential $\mu$ and mass $m$ is related to the weight at $\mu=0$ and effective mass $m/\sqrt{1-\mu^2}$. 
Note that for $\mu>1$ the effective mass on the right-hand side becomes imaginary, and the subset weights, although still real, can be either positive or negative, without having a definite sign. Therefore, importance sampling of subsets can only be used for $\mu < 1$, which is the relevant region when relating the random matrix model to QCD.

\subsection{$\mathbf{N_f}$ massive fermions}

The proof in the previous section can be generalized to an arbitrary number of degenerate flavors, in exactly the same way as was done in Sec.~\ref{sec:Nfneq1} for the massless case. Equation~\eqref{conjectmassive} is then replaced by
\begin{align}
\lefteqn{\ssum {\det}^{N_f} Q_\theta(\mu,m)} \notag\\
&= (1-\mu^2)^{N_f N} \ssum {\det}^{N_f} Q_\theta\left(0,\frac{m}{\sqrt{1-\mu^2}}\right) ,
\end{align}
and after multiplying with $m^{N_f \nu}$ and using Eq.~\eqref{detDmum}, we finally find
\begin{align}
\lefteqn{\ssum {\det}^{N_f} D_\theta(\mu,m)} \\
&= (1-\mu^2)^{N_f (N+\nu/2)} \ssum {\det}^{N_f} D_\theta\left(0,\frac{m}{\sqrt{1-\mu^2}}\right) , \notag
\end{align}
which proves the conjecture \eqref{conjectmneq0} for arbitrary $N_f$.

\section{Implementation}
\label{sec:implementation}

Below we describe the numerical implementation of the computation of the determinant, chiral condensate and quark number density.
We write the Dirac matrix \eqref{Ddef} as
\begin{align}
D =
  \begin{pmatrix}
    m_{N+\nu} & A \\
    B & m_N
  \end{pmatrix}\:,
\end{align}
where $A = \imath\phi_1 + \mu\phi_2$ is an $(N+\nu) \times N$ matrix, $B = \imath\phi_1^\dagger + \mu\phi_2^\dagger$ is $N \times (N+\nu)$ and $m_N = \diag(m \ldots m)$ is a diagonal $N \times N$ matrix.

\paragraph{Determinant.}

It is well-known that the determinant of such a block matrix can be computed as
\begin{align}
\det D &= \det\left[m_{N+\nu}\right] \det\left[m_N - B m_{N+\nu}^{-1} A\right]
\end{align}
when $m \neq 0$.
Note that this product of determinants cannot be merged as their arguments have different dimensions. This can be simplified to 
\begin{align}
\det D = m^{\nu} \det Q ,
\label{mdet}
\end{align}
where we defined 
\begin{align}
Q \equiv m^2_N - BA .
\label{Q}
\end{align}
Note that the factor $m^\nu$ is reminiscent of the $\nu$ exact zero modes of the massless Dirac matrix. In the massless case the determinant simplifies to 
\begin{align}
\det D = \det [-BA] ,
\label{mdet0}
\end{align}
for $\nu=0$ (or for $\nu \neq 0$ if we neglect the zero modes).
Numerically, $\det Q$ was computed using an LU-factorization, as this is more efficient and accurate than using a full diagonalization.

\paragraph{Chiral condensate.}

The chiral condensate is given by (for $N_f=1$)
\begin{align}
\Sigma &= \frac{1}{2N}\frac{\partial \log Z}{\partial m}
= \frac{1}{2N}\frac{1}{Z} \Dphi \frac{\partial \det D}{\partial m} \notag\\
&= \frac{1}{2N}\frac{1}{Z} \Dphi \, \det D \,
 \tr\left[\frac{\partial D}{\partial m}  \, D^{-1} \right] \notag\\
&= \frac{1}{2N}\frac{1}{Z} \Dphi \, \det D \, \tr D^{-1} \notag\\
&= \left\langle\frac{1}{2N} \tr D^{-1} \right\rangle ,
\label{Strace}
\end{align}
where we denoted $\Dphi\equiv\int d\phi_1 d\phi_2 \, w(\phi_1) \, w(\phi_2)$.
The inverse of $D$ can be computed efficiently using its block structure, which yields
\begin{align}
D^{-1} &=
  \begin{pmatrix}
    \frac{1}{m} \left(\one_{N+\nu} + A Q^{-1} B\right) & \quad - A Q^{-1} \\
    - Q^{-1} B & m Q^{-1}
  \end{pmatrix}
.
\label{Dinv}
\end{align}
This formula is easily verified as $D D^{-1} = 1$.
When taking the trace of \eq{Dinv} to compute \eqref{Strace} this can be further simplified as
\begin{align}
\Sigma &= \frac{1}{2N}\left\langle \tr\left[ \frac{1}{m} \left(\one_{N+\nu} + A Q^{-1} B\right) \right] + \tr\left[ m Q^{-1} \right] \right\rangle \notag\\
&= \frac{1}{2Nm} \left\langle N+\nu + \tr\left[ (m^2 + B A) Q^{-1} \right] \right\rangle \notag\\
&= \frac{1}{2Nm} \left\langle N+\nu + \tr\left[ (2m^2 - Q) Q^{-1} \right] \right\rangle\notag\\
&= \frac{\nu}{2Nm} + \frac{m}{N}\langle \tr Q^{-1} \rangle
,
\label{Simplem}
\end{align}
where we also used the definition \eqref{Q} of $Q$. The term $\nu/2Nm$ originates from the zero modes of the massless operator and is sometimes omitted.
The inverse $Q^{-1}$ can be computed using the LU-factorization of $Q$, which is already available from the evaluation of $\det Q$.

\paragraph{Quark number density.}

The average quark number density is given by (for $N_f=1$)
\begin{align}
n &= \frac{1}{2N}\frac{\partial \log Z}{\partial \mu}
= \frac{1}{2N}\frac{1}{Z} \Dphi
  \frac{\partial \det D}{\partial \mu} \notag\\
&= \frac{1}{2N}\frac{1}{Z} \Dphi \, \det D \,
 \tr\left[\frac{\partial D}{\partial \mu}  \, D^{-1} \right] \notag\\
&= \frac{1}{2N}\frac{1}{Z} \Dphi  \, \det D \,
 \tr\left[\begin{pmatrix}
0 & \phi_2 \\
\phi_2^\dagger & 0
\end{pmatrix} 
D^{-1} \right] \notag\\
&= \left\langle\frac{1}{2N} \tr \left[
\begin{pmatrix}
0 & \phi_2 \\
\phi_2^\dagger & 0
\end{pmatrix}
D^{-1}
\right] \right\rangle
  .
\label{qn}
\end{align}
Using the block-inverse \eqref{Dinv} this simplifies to
\begin{align}
n &= 
 -\frac{1}{2N} \left\langle \tr\left[ (\phi_2^\dagger A + B\phi_2) Q^{-1} \right] \right\rangle.
\label{nimplem}
\end{align}

\section{Reweighting}
\label{sec:rew}

Reweighting methods can be used to perform Monte Carlo simulations when the weights are complex and the ensemble cannot be directly sampled with importance sampling.

The ensemble average of an observable $y(x)$ in an ensemble with weights $w(x)$ is defined by
\begin{align}
\langle {y} \rangle_{w} = \frac{\int dx \; w(x) y(x)}{\int dx \; w(x)} .
\end{align}
In the reweighting method one introduces an \textit{auxiliary} ensemble with weights $\waux(x)$ and rewrites the previous equation as
\begin{align}
\langle {y} \rangle_{w} = \frac{\int dx \; \waux(x) \frac{w(x)}{\waux(x)} y(x)}{\int dx \; \waux(x) \frac{w(x)}{\waux(x)}} 
= \frac{\left\langle {\frac{w}{\waux} y} \right\rangle_{\!\waux}}{\left\langle {\frac{w}{\waux}} \right\rangle_{\!\waux}} .
\label{reweight}
\end{align}
If the weights $\waux$ are chosen to be real and positive, the auxiliary ensemble can be sampled using importance sampling methods and the ratio of expectation values in Eq.~\eqref{reweight} can be evaluated in a Monte Carlo simulation.
Typical examples for $\waux$ are the quenched, phase-quenched, $\mu$-quenched and sign-quenched ensembles.

Reweighting methods typically suffer from an overlap problem, when the relevant configurations in the target and auxiliary ensembles do not coincide. More importantly, when the target weight is non-positive one encounters the sign problem, as the work needed to make reliable measurements on the statistical ensemble grows exponentially with volume and chemical potential because it involves the computation of exponentially small reweighting factors $\left\langle w/\waux \right\rangle_{\!\waux}$ from a statistical sampling of largely canceling contributions \cite{deForcrand:2010ys}.

\section{Some analytical results}

In order to verify some of our simulation data, we quote a couple of known analytical results.

\subsection{$\mathbf{N_f=1}$ observables}
\label{sec:Nf=1}

The $N_f=1$ partition function $Z_1$ can be expressed in terms of orthogonal polynomials as \cite{Osborn:2008jp,Akemann:2002vy}
\begin{align}
\frac{Z_{1}(\mu;m)}{Z_{0}} = m^\nu \left(\frac{1-\mu^2}{N}\right)^N N! \, L_N^\nu\left(-\frac{N m^2}{1-\mu^2}\right) ,
\end{align}
where $L_N^\nu$ are generalized Laguerre polynomials of order $\nu$ and degree $N$.
Their derivatives are given by
\begin{align}
\frac{d}{dz} L_N^\nu(z) = - L_{N-1}^{\nu+1}(z),
\end{align}
such that the chiral condensate \eqref{ccdef} is given by \cite{Osborn:2008jp}
\begin{align}
\Sigma(\mu;m) &= \frac{\nu}{2Nm} + \frac{m}{1-\mu^2}
 \frac{L_{N-1}^{\nu+1} \left(-\frac{N m^2}{1-\mu^2}\right) }{ L_N^\nu\left(-\frac{N m^2}{1-\mu^2}\right) } .
\label{Stheo}
\end{align}
The quark number density \eqref{qndef} can be computed analogously, yielding
\begin{align}
n(\mu;m) &= -\frac{\mu}{1-\mu^2} \left[ 1 -   \frac{m^2}{1-\mu^2} 
\frac{L_{N-1}^{\nu+1}\left(-\frac{N m^2}{1-\mu^2}\right)}{L_N^\nu\left(-\frac{N m^2}{1-\mu^2}\right)} 
\right] .
\label{ntheo}
\end{align}

In the microscopic limit, where $\hat m = 2N m$ and $\hat\mu^2=2N\mu^2$ are kept fixed while taking $N \to\infty$ the chiral condensate \eqref{Stheo} and quark number density \eqref{ntheo} become
\begin{align}
\hat\Sigma(\hat\mu;\hat m) = \frac{I_\nu'(\hat m)}{I_\nu(\hat m)} \quad \text{and} \quad
\hat n(\hat\mu;\hat m) = 0 ,
\label{theomiclim}
\end{align}
where $I_\nu$ is a modified Bessel function.

\subsection{Phase-quenched reweighting factor for $\mathbf{N_f=2}$}
\label{sec:averagephase}

The reweighting factor, i.e.\ the denominator in the reweighting formula \eqref{reweight}, in the phase-quenched reweighting scheme is nothing but the average phase of the fermion determinant in the phase-quenched ensemble and can be computed analytically for $N_f=2$. This average phase can be written as \cite{Splittorff:2007ck}
\begin{align}
\langle e^{2\imath \theta} \rangle_\text{pq} 
= \frac{\langle {\det}^2 D \rangle_{N_f=0}}{\langle |{\det} D|^2 \rangle_{N_f=0}} .
\label{avgphase}
\end{align}
Both, numerator and denominator are quenched expectation values of products of characteristic polynomials and their complex conjugates, which can be computed analytically using the method of orthogonal polynomials \cite{Akemann:2002vy}. 
For random matrices of size $N$ and topology $\nu$ this yields,
\begin{align}
\langle {\det}^2 D \rangle_{N_f=0}
&= \frac{1}{2m} \det 
\begin{pmatrix}
p_N^\nu(m;\mu) & p_{N+1}^\nu(m;\mu) \\[2mm]
\partial_m p_N^\nu(m;\mu) & \partial_m p_{N+1}^\nu(m;\mu) \\
\end{pmatrix}
\label{Z2}
\end{align}
and
\begin{align}
\langle |{\det} D|^2 \rangle_{N_f=0}
&= r_N^\nu(\mu) \sum_{k=0}^N \frac{\left|p_k^\nu(m;\mu)\right|^2}{r_k^\nu(\mu)}  ,
\label{Z11*}
\end{align}
with orthogonal polynomials
\begin{align}
p_k^\nu(z;\mu) = \left(\frac{1-\mu^2}{N}\right)^k k! \, L_k^\nu\left(-\frac{N z^2}{1-\mu^2}\right) 
\end{align}
 and normalization factors
\begin{align}
r_k^\nu(\mu) = \frac{1}{N^{2k+\nu+2} }\pi \mu^2 (1+\mu^2)^{2k+\nu} k! (k+\nu)! \,.
\end{align}
Substitution of Eqs.~\eqref{Z2} and \eqref{Z11*} in \eq{avgphase} yields the average phase in the phase-quenched ensemble.

\newpage

\section{Silver Blaze at finite $N$}
\label{app:gamma}

Below we show that the partition function \eqref{partfun} can be made independent of $\mu$ with a judicious choice of $\fac$ in the Gaussian weights \eqref{rmtdis}. For arbitrary $\gamma \in \mathbb{R}^+$ we rescale the matrices in \eqref{partfun} using $\phi_i'=\sqrt{\fac}\phi_i$, such that
\begin{align}
  Z = C \int & d\phi_1' d\phi_2' \,
  \exp[-N (\tr {\phi_1'}^\dagger \phi_1'+\tr {\phi_2'}^\dagger \phi_2')] \notag \\
  & \times 
  \prod_{f=1}^{N_f} {\det} \, D(\phi_1'/\sqrt{\fac}, \phi_2'/\sqrt{\fac};\mu,m_f) ,
\end{align}
where $C=\left(N/\pi\right)^{2N(N+\nu)}$ and we also took into account the Jacobian of the transformation. The $\fac$-dependence has thus been shifted from the Gaussian weights to the Dirac matrix, and the partition function now looks like a conventional $\gamma=1$ partition function, albeit with a modified Dirac matrix. From the structure of the Dirac matrix \eqref{Ddef} we see that the scaling of the fields can be shifted to the mass, as 
\begin{align}
D(\phi_1'/\sqrt{\fac}, \phi_2'/\sqrt{\fac};\mu,m_f) = \frac{1}{\sqrt{\fac}} D(\phi_1', \phi_2';\mu,\sqrt{\fac}\, m) .
\end{align}
Recalling that the Dirac matrix has dimension $2N+\nu$, the partition function becomes,
\begin{align}
Z = & C \int d\phi_1' d\phi_2' \,
\exp[-N (\tr {\phi_1'}^\dagger \phi_1'+\tr {\phi_2'}^\dagger \phi_2')] \notag \\
  & \times 
  \,\gamma^{-N_f(2N+\nu)/2}\prod_{f=1}^{N_f} {\det} \, D(\phi_1', \phi_2';\mu,\sqrt{\fac}\, m_f) ,
\label{Zgamma}
\end{align}
We now look at the subset sum \eqref{sigma} of fermion determinants for $N_f$ degenerate quarks in the partition function \eqref{Zgamma}. Using the relation \eqref{conjectmneq0} we find that these fermionic subset weights satisfy
\begin{align}
\lefteqn{\gamma^{-N_f(2N+\nu)/2} \, \sigma_\Omega(\mu,\sqrt{\fac} m)  } \notag\\[1mm]
& \hspace{4ex} = \left(\frac{1-\mu^2}{\gamma}\right)^{N_f(N+\nu/2)}\,\Sdet_\Omega\left(0,\sqrt{\frac{\gamma}{1-\mu^2}} \, m\right) .
\label{sigmagamma}
\end{align}
Clearly, if we choose $\gamma=1-\mu^2$ (assuming $\mu\leq 1$)  the right-hand side of \eqref{sigmagamma} is independent of $\mu$. From Eq.~\eqref{Zgamma} we can then immediately conclude that Eq.~\eqref{Zrel} is now  replaced by
\begin{align}
Z_{N_f}(\mu;m) &= Z_{N_f}\left(0;m\right) ,
\label{Zrelgamma}
\end{align}
such that the Silver Blaze is not only satisfied in the microscopic limit, but also for any finite $N$ away from it.

\cleardoublepage

\bibliography{biblio} 

\end{document}